\documentclass[aps,prd,showpacs,preprintnumbers]{revtex4-1}
\usepackage{epsfig}
\usepackage{latexsym,amsmath,amssymb,amstext}
\usepackage{float}
\usepackage{graphicx,xcolor}

\newcommand{\be}{\begin{equation}}
\newcommand{\ee}{\end{equation}}
\newcommand{\bea}{\begin{eqnarray}}
\newcommand{\eea}{\end{eqnarray}}

\newcommand{\feyn}[1]{#1\!\!\!\!\slash\  }
\newcommand{\tc}{{\cal T}}
\newcommand{\pac}{{\cal F}}
\newcommand{\pbc}{{\cal S}}
\newcommand{\patc}{\widetilde{\cal F}}
\newcommand{\pbtc}{\widetilde{\cal S}}
\newcommand{\rc}{{\cal R}}

\newcommand{\cc}{{\cal C}}
\newcommand{\hc}{{\cal H}}

\newcommand{\ket}[1]{{|#1\rangle}}

\newcommand\bef{\begin{figure}}
\newcommand\eef[1]{\label{fg:#1}\end{figure}}
\newcommand\beq{\begin{equation}}
\newcommand\eeq[1]{\label{#1}\end{equation}}
\newcommand\beqa{\begin{eqnarray}}
\newcommand\eeqa[1]{\label{#1}\end{eqnarray}}
\newcommand\bet{\begin{table}}
\newcommand\eet[1]{\label{tb:#1}\end{table}}

\newcommand\fgn[1]{Figure \ref{fg:#1}}
\newcommand\eqn[1]{Eq.\ (\ref{#1})}
\newcommand\scn[1]{Section \ref{sec:#1}}
\newcommand\apx[1]{Appendix \ref{sec:#1}}
\newcommand\tbn[1]{Table \ref{tb:#1}}

\newcommand\ie{{\sl i.e.\/}}

\newcommand{\st}{{\rm st}}

\begin{document}
\title{Phase of the fermion determinant in QED$_3$ using a gauge invariant lattice regularization}
\author{Nikhil\ \surname{Karthik}}
\email{nkarthik@fiu.edu}
\affiliation{Department of Physics, Florida International University, Miami, FL 33199.}
\author{Rajamani\ \surname{Narayanan}}
\email{narayanr@fiu.edu}
\affiliation{Department of Physics, Florida International University, Miami, FL 33199.}

\begin{abstract}
We use canonical formalism to study the fermion determinant in different
three dimensional abelian gauge field backgrounds  that contain non-zero
magnetic and electric flux in order to understand the non-perturbative
contributions to the parity-odd and parity-even parts of the phase.  We
show that a certain phase associated with free fermion propagation along
a closed path in a momentum torus is responsible for the parity anomaly
in a background with non-zero electric flux. We consider perturbations
around backgrounds with non-zero magnetic flux to understand the structure
of the parity-breaking perturbative term at finite temperature and mass.
\end{abstract}

\date{\today}

\maketitle

\section{Introduction}

Three dimensional Euclidean abelian gauge theory coupled to a two
component massless fermion  by
\be
\feyn{D}({\bf A}) = \sigma_k \left( \partial_k + i A_k\right) ,
\ee
can induce a parity breaking mass term for the gauge field in the form
of a Chern-Simons action~\cite{Deser:1981wh,Deser:1982vy,Niemi:1983rq}
\be
S = \frac{i\kappa}{4\pi} \int d^3 x A_k F_k\qquad \text{where}\qquad
F_k = \epsilon_{kij} \partial_i A_j. \label{cs}
\ee
The mass term is gauge invariant
in infinite volume provided the fields are assumed to vanish at infinity.
This result has been shown in perturbation theory using the gauge invariant
Pauli-Villars regularization~\cite{Redlich:1983dv}. 

Theories with $2N$ flavors of massless fermions can have a real and
positive determinant with proper pairing of fermions. Vacuum energy
arguments show that the $O(2N)$ symmetry breaks down to $O(N) \times O(N)$
symmetry~\cite{Vafa:1984xh} and it has been shown that dynamical masses
are generated for fermions that do not break parity~\cite{Pisarski:1984dj}
in the large $N$ limit. Recent calculations along the same lines using
Schwinger-Dyson equations~\cite{Braun:2014wja} attempt to identify
the phase structure that separates one where dynamical masses are
generated from others that do not.  Numerical studies using staggered
fermions~\cite{Hands:2002dv,Hands:2004bh,Fiore:2005ps} have been performed
and condensates have been computed for theories that do not break parity,
again with the aim of exploring the phase structure.

The gauge theory with one flavor of two component Dirac fermion
can be regularized in a gauge invariant manner using the Wilson-Dirac
operator~\cite{So:1984nf, So:1985wv, Coste:1989wf}
\be
D_w({\bf U},M) = \feyn{D}_n({\bf U}) - B({\bf U}) + M ;\qquad
\feyn{D}_n^\dagger({\bf U}) = - \feyn{D}_n({\bf U});\qquad B^\dagger({\bf U}) = B({\bf U}),\label{wilsonp}
\ee
where ${\bf U}$ is the U(1) valued lattice link variable; $\feyn{D}_n({\bf
U})$ is the na\"ive lattice fermion operator; $B({\bf U})$ is the Wilson
term that provides a mass of the order of cut-off for the doublers;
and $0 \le M < 1$ is the mass in lattice units for the fermion.
Perturbation theory computations~\cite{Coste:1989wf} using \eqn{wilsonp}
in infinite volume show that the coefficient of the induced Chern-Simons
term in \eqn{cs} is $\kappa=-1$ if $M>0$, and $\kappa=-\frac{1}{2}$
if one takes the massless limit after taking the continuum limit.
For negative fermion masses, we would use
\be
D_w({\bf U},-M) = -D^\dagger_w({\bf U},M) = \feyn{D}_n({\bf U}) + B({\bf U}) - M\qquad\text{for}\qquad 0 \le M < 1,\label{wilsonn}
\ee
so that the induced parity breaking term as one approaches the massless
limit from the positive and negative side are opposite in sign.

A theory with $2N$ flavors with $N$ flavors obeying \eqn{wilsonp}
and the other $N$ flavors obeying \eqn{wilsonn} can be used for
a numerical investigation of condensates that do not break parity.
We can also consider theories with non-degenerate fermions and arbitrary
number of flavors, and study the effect of parity breaking mass terms in
the limit of large number of flavors.

Consider the continuum limit in a lattice simulation where we take the
number of lattice points denoted by $L\to\infty$. The continuum limit needs to be taken keeping the 
physical spatial extent $l$, the fermion mass $m_{\rm phys}$ and the temperature $T$ constant as $L\to\infty$. 
In a lattice calculation, it is natural to instead consider
the dimensionless temperature $t=l T$ and the dimensionless mass $m= l m_{\rm phys}$, measured in units of
the spatial extent, to be the parameters of the theory and keep them constant as $L\to\infty$. 
Since we study fermions on fixed 
gauge field backgrounds, the coupling constant $g^2$ does not play a role in the present calculations. 
The induced gauge action in a fixed gauge field background will be gauge invariant and it
is of interest to study this outside perturbation theory before embarking
on a full lattice simulation. Of particular interest is the phase of
fermion determinant which contains parity violating terms. Consider,
for example~\cite{Dunne:1998qy}, a gauge field background that has a
non-zero magnetic flux,
\be
\int F_3 dx dy = 2\pi q_3,
\ee
for integer $q_3$, along with a non-zero Polyakov loop, $e^{i\int A_3
d \tau} = e^{i2\pi h_3}$. The associated Chern-Simons action is
\be
S(h_3, q_3) = i \kappa \pi h_3 q_3,
\ee
and it has to remain invariant under the gauge transformation
$h_3 \to (h_3+1)$.  This implies $\kappa$ has to be an even
integer~\cite{Pisarski:1986gr,Henneaux:1986tt,Hosotani:1989vz}
for this particular gauge field background in a regularization
that preserves gauge invariance  under such ``large" gauge
transformations~\cite{Deser:1997nv,Deser:1997gp,Fosco:1997ei}. This
does not match with $\kappa=-1$ or $\kappa=-\frac{1}{2}$ obtained
in~\cite{Coste:1989wf}. 
Effect of non-vanishing gauge fields at infinity on spontaneous and anomalous breaking of 
parity have also been addressed in~\cite{Nissimov:1985wk}.
In addition to the parity violating contributions
to the phase of the fermion determinant,
\be
e^{i \Gamma(t,m,{\bf A})} = \lim_{L\to\infty} \frac{ \det D_w(U,M)}{\left| \det D_w(U,M)\right|},
\ee
in the continuum limit, there are also parity preserving
 contributions of the form $e^{i\pi h(A)}$ where $h(A)$ are
 integers associated with zero crossings of the Wilson-Dirac
operator~\cite{AlvarezGaume:1984nf,Forte:1986em,Forte:1987kj}.

As a precursor to studying the three dimensional theory,
consider the regularized result using Wilson-Dirac fermions
in one dimension in comparison to the results obtained
in~\cite{Dunne:1998qy,Deser:1997nv,Deser:1997gp,Fosco:1997ei,Kikukawa:1997qh}.
The Wilson-Dirac fermion operator in one dimension is
\be
D_w(U,M) = -1 + M + T,
\ee
where the translation operator $T$ is $(T\psi)(k) = U(k)\psi(k)$ in terms
of the one dimensional link variable, $U(k)$.  The only physical degree
of freedom is the Polyakov loop,
\be
W = \lim_{L\to\infty}\prod_{k=1}^L U(k)= e^{i2\pi h},
\ee
and the fermion determinant in the continuum, assuming $L$ to be even, is
\be
\lim_{L\to\infty} \det D_w\left(U,\frac{m}{L}\right) = 
\begin{cases}
 e^{ i2\pi h} - e^{-m}=
\begin{cases}
e^{i2\pi h } & m=\infty \cr
2\sin (\pi h) e^{i\pi h + i\frac{\pi}{2}} & m=0_+
\end{cases} \cr
e^{ -i2\pi h} - e^{-m}=
\begin{cases}
2\sin (\pi h) e^{-i\pi h - i\frac{\pi}{2}} & m=0_-\cr
e^{-i2\pi h} & m=-\infty .
\end{cases}
\end{cases}
\ee
The result matches with the one obtained in~\cite{Deser:1997gp} using
zeta function regularization and has the main features discussed before.
It is invariant under the ``large" gauge transformation $h\to (h+1)$ for
all values of $m$. The part of phase proportional to $h$ in the massless
limit is half of its value in the infinite mass limit.  As for the vacuum
structure is concerned, the partition function for a two flavor theory
with masses $m_1$ and $m_2$ is
\be
Z(m_1,m_2) = \begin{cases}
e^{-(m_1+m_2)} & m_1,m_2 > 0\cr
e^{-(m_1-m_2)}+1 & m_1>0;\ \  m_2<0,
\end{cases}
\ee
showing that the theory with $m_1>0$ and $m_2 <0$ is preferred over
$m_1,m_2 > 0$.

The aim of this paper is to study the phase $\Gamma(t,m,{\bf A})$ in the
continuum U(1) gauge field background  on a three dimensional $l\times
l \times \frac{1}{t}$ torus given by
\be
A_1 = \frac{2\pi q_2 t}{l} \tau + \frac{2\pi h_1}{l} + A_1^p; \qquad
A_2 = \frac{2\pi q_3}{l^2} x + \frac{2\pi h_2}{l} + A_2^p; \qquad
A_3 = \frac{2\pi q_1 t}{l} y + 2\pi h_3 t+ A_3^p ,
\label{contfield}
\ee
where $q_i$ are integers and they denote non-zero flux in the $x$, $y$ and
$\tau$-directions; $h_i\in [0,1]$ denotes torons generating non-trivial
Polyakov loops; and $A_i^p$ are perturbative fields that obey periodic
boundary conditions.  The associated {\sl periodic} boundary conditions
on fermions are
\be
\psi(l,y,\tau) = e^{-i\frac{2\pi q_3 y}{l}} \psi(0,y,\tau);\qquad
\psi(x,l,\tau) = e^{-i2\pi q_1 \tau t} \psi(x,0,\tau);\qquad
\psi(x,y,\frac{1}{t}) = e^{-i\frac{2\pi q_2 x}{l}} \psi(x,y,0).\label{fermbound}
\ee
We refer to $q_3$ as magnetic flux, and $q_1$ and $q_2$ as electric flux.
The naming is not relevant if only one of the three is non-zero,
but we also consider cases with $q_1\ne 0$, $q_2\ne 0$ and $q_3=0$
in this paper and extract some results without completely resorting to
numerical means. In addition, we numerically study the most general case
with non-zero flux in all three directions.  The phase splits into a
parity even and odd part
\be
\Gamma = \Gamma_{\rm even}+\Gamma_{\rm odd},
\ee
with the parity even part being
\be
\Gamma_{\rm even}=\pi (q_1 + q_2 + q_3) +\pi( q_1q_2 + q_3q_1 + q_2q_3).
\label{evenintro}
\ee
The first term can be absorbed by changing the boundary conditions
of fermions but not both the first and second terms.  In general, the
parity odd part is complicated, but it has a simple form in the case of
zero temperature when we consider a $\tau$-dependent perturbation on a
static and spatially uniform magnetic field:
\be
\Gamma_{\rm odd} =-2\pi h_3 q_3 -\int d\tau d \tau'  A_1^p(\tau) A_2^p(\tau') G(\tau-\tau'),
\ee
where the form factor $G(\tau)$ is an odd function of $\tau$ that depends
on the fermion mass $m$ and spatial torons.  Our formulation on the
lattice enables us to study $G(\tau)$ without making prior assumptions
concerning the local or non-local nature of the induced gauge action. We
study how the form factor becomes local in the limit of $m\to\infty$
and $m\to 0$.

The organization of the paper is as follows. We describe lattice
gauge fields on a torus in \scn{torus}.  In \scn{canform}, we derive
an expression for the Wilson-Dirac fermion determinant in the lattice
axial gauge allowing for non-trivial Polyakov loops using the canonical
formalism~\cite{Hasenfratz:1991ax}.  In \scn{em}, we use the canonical
formalism to study cases with uniform electric and magnetic fields,
 organized into subsections. Here, we explain the origin of the parity
even phase in \eqn{evenintro}.  First, we present a conventional way to
understand the parity breaking when there is only a non-zero magnetic
flux.  The  zero crossings of the eigenvalues of the two dimensional Dirac
operator are responsible for the parity breaking terms and the formula
for the fermion determinant using lattice regularization matches the
one from zeta function regularization~\cite{Deser:1997nv,Deser:1997gp}.
We then consider the case where we have non-zero electric fluxes but
zero magnetic flux.  We show that the relevant quantity to obtain the
parity even part of the phase is associated with the propagation of a
free fermion with continuously changing momentum along a closed loop
in the torus in momentum space in a direction defined by $(q_2,-q_1)$.
Finally, we turn on perturbations over static magnetic field backgrounds.
For this, we develop second order perturbation theory with in the canonical
formalism in \scn{pert} and use it to study the parity odd part of the
induced effective action. The results of the perturbative analysis and the
numerical extraction of the form factor $G$ are presented in \scn{pert0}
and \scn{pertemp}.

\section{Gauge field on a torus}\label{sec:torus}

We work on an $L^2\times\beta$ lattice for the sake of simplicity,
 which can be easily generalized to a spatially  anisotropic lattice
as well. We only consider lattices where both $L$ and $\beta$ are even;
while the  continuum physics is independent of this choice, it helps
to simplify our  calculations. The spatial volume of the lattice is
defined as $V\equiv L^2$. The spatial lattice points are labelled by
$\mathbf{x}=(x_1,x_2)$ with $ 1 \le x_i \le L$, and the temporal lattice
points by $k$ with $1\le k \le \beta$.  The dimensionless temperature
in the continuum limit is
\be
t=\lim_{L\to\infty} \frac{L}{\beta}.
\ee
The continuum space and Euclidean time variables are 
\be
(x,y)=\lim_{L\to\infty} \left( \frac{x_1}{L},\frac{x_2}{L}\right)\qquad\text{and}\qquad
\tau=\lim_{L\to\infty} \frac{k}{L},
\ee
with $x,y\in [0,1]$ and $\tau\in \left[ 0,\frac{1}{t}\right]$.

On this lattice, we introduce U$(1)$ gauge fields using the gauge-links
$U_\mu(\mathbf{x},k)$. In this work, we fix the gauge such that the
temporal gauge-links from $k=1$ to $k=\beta-1$ are set to identity.
Non-trivial Polyakov loop variables in the $\tau$-direction are taken
care of by the presence of $U_3(\mathbf{x},\beta)=U_3(\mathbf{x})$. This
partial gauge fixing enables us to develop the canonical formalism in
\scn{canform}. We still have a remnant time independent gauge symmetry,
$g(\mathbf{x})$, under which
\be
U_i(\mathbf{x},k) \to g^\dagger(\mathbf{x}) U_i(\mathbf{x},k) g(\mathbf{x}+\hat i)\quad\text{and}\quad U_3(\mathbf{x}) \to g^\dagger(\mathbf{x}) U_3(\mathbf{x}) g(\mathbf{x}).
\ee
In this work, we consider only the gauge fields of the form in
\eqn{contfield}. An analogous Hodge-decomposition is strictly true for any
gauge-fields in two dimensions. In three dimensions, one should consider
these as specific background  gauge fields used in order to probe the
dependence of the fermion determinant on perturbative and non-perturbative
aspects  of the gauge-field.  The gauge fields in \eqn{contfield} are
periodic only up to a gauge transformation with non-trivial winding. Since
we do not require smoothness of the link variables on the lattice, such
gauge fields along with fermions, which satisfy the boundary conditions in
\eqn{fermbound}, can be incorporated  using gauge links and fermions that
are strictly periodic.  For our gauge choice, the lattice gauge field
background corresponding to \eqn{contfield} is
\bea
U_1({\bf x}, k) &=& \begin{cases}
e^{i\frac{2\pi q_2}{L \beta}k}e^{i \frac{2\pi h_1}{L}+i A_1^q} & \text{if}\quad x_1 < L \cr
e^{i\frac{2\pi q_2}{L \beta}k-i\frac{2\pi q_3}{L}x_2}e^{i\frac{2\pi h_1}{L}+i A_1^q} &\text{if}\quad x_1 = L, \cr
\end{cases}\cr
U_2({\bf x}, k) &=& 
e^{i\frac{2\pi q_3}{L^2}x_1-i\frac{2\pi q_1}{\beta L}k}e^{i\frac{2\pi h_2}{L}+i A_2^q}, \cr
U_3({\bf x}) &=& 
e^{-i\frac{2\pi q_2}{L}x_1+i\frac{2\pi q_1}{L}x_2}e^{i 2\pi h_3}.
\label{gfield}
\eea
The various background gauge fields we study in this paper are instances
of the above equation.

\section{Canonical formalism}\label{sec:canform}

The partial gauge fixing defined in \scn{torus} naturally
allows for the development of the Hamiltonian or the canonical
formalism~\cite{Hasenfratz:1991ax}.  Let $T_i(k)$ be the parallel
transporters along spatial directions at a fixed Euclidean time, $k$:
\be
\left[T_j(k) \psi\right](\mathbf{x})\equiv U_j(\mathbf{x},k) \psi(\mathbf{x}+\hat j),
\ee
and let $T_3$ defined as
\be
\left[T_3 \psi\right](\mathbf{x})\equiv U_3(\mathbf{x}) \psi(\mathbf{x}),
\ee
be the parallel transporter that connects $k=\beta$ and $k=1$.
In this gauge field background, the Wilson-Dirac fermion operator is
\bea
D^{kk'}(M) &=& \left[-3 +M
+\frac{1}{2} \sum_{i=1}^2\left[ \left(\sigma_i+1\right) 
  T_i(k)
-\left (\sigma_i-1\right) T_i^\dagger(k)\right] \right] \delta^{k',k}\cr
&&+\frac{1}{2}
\begin{cases}
\left[   \left( \sigma_3+1\right)
  \delta^{k',2} -\left (\sigma_3-1\right) T_3^\dagger \delta^{k',\beta} \right] &\text{if}\quad k=1 \cr
 \left[ \left(\sigma_3+1\right) 
  \delta^{k',k+1}
-\left (\sigma_3-1\right)\delta^{k',k-1}\right] &\text{if}\quad 1 < k < \beta\cr
\left[ \left(\sigma_3+1\right)
  T_3 \delta^{k',1} -\left (\sigma_3-1\right) \delta^{k',\beta-1}\right] &\text{if}\quad k= \beta, \cr
\end{cases}
\label{wfer}
\eea
where the second term takes care of the periodicity in the temporal
direction. In this way, we have managed to write the Dirac operator using
operators defined on two-dimensional time-slices. The Wilson mass $M$
is such that $|M| < 1$. It is related to the physical mass $m$ in the
units of box length as
\be
m\equiv M L.
\ee
By using the following set of Pauli matrices,
\be
\sigma_1 = \begin{pmatrix} 0 & 1 \cr 1 & 0 \cr \end{pmatrix};\ \ \ \ 
\sigma_2 = \begin{pmatrix} 0 & -i \cr i & 0 \cr \end{pmatrix};\ \ \ \ 
\sigma_3 = \begin{pmatrix} 1 & 0 \cr 0 & -1\cr \end{pmatrix},
\ee
the Wilson-Dirac operator $D$ can be written in the matrix form as 
\be 
D(M)= \begin{pmatrix}
-B_1 & C_1 & 1 & 0 & \cdots & \cdots & 0& 0 \cr 
-C_1^\dagger  & -B_1 & 0 & 0& \cdots & \cdots & 0 & T_3^\dagger \cr 
0 & 0 & -B_2 & C_2 & \cdots & \cdots & 0 & 0 \cr 
0 & 1 & -C_2^\dagger & -B_2 &  \cdots & \cdots & 0 & 0 \cr 
\vdots & \vdots & \vdots & \ddots & \ddots & \ddots & \vdots & \vdots \cr 
\vdots & \vdots & \vdots & \vdots & \ddots & \ddots & \ddots & \vdots \cr 
T_3 & 0 & 0 & 0 & \cdots & \cdots & -B_{\beta} & C_{\beta} \cr 
0 & 0 & 0 & 0 & \cdots & \cdots & -C_{\beta}^\dagger & -B_{\beta} \cr 
\end{pmatrix},
\ee
where
\bea
B_k &\equiv& 3 - M - \frac{1}{2} \sum_{j=1}^2 \left(T_j(k) + T^\dagger_j(k)\right),\cr 
C_k &\equiv& \frac{1}{2} \left(T_1(k) -
  T^\dagger_1(k)\right)-\frac{i}{2}\left(T_2(k) -
  T^\dagger_2(k)\right).
\eea 
Note that $B_k$ is a positive definite operator for $|M| < 1$. We closely follow~\cite{Neuberger:1997bg}
in order to obtain an expression for the determinant of $D$. We first
cyclically permute the columns to the left. This gives a matrix
\be
D'(M)= \begin{pmatrix}
\alpha_1 & 0 & 0 & \cdots & \cdots & 0 & \gamma_1 Y \cr 
\gamma_2 & \alpha_2 & 0 & \cdots & \cdots & 0 & 0 \cr 
0 & \gamma_3 & \alpha_3 & \cdots & \cdots & 0 & 0 \cr 
\vdots & \vdots & \vdots & \ddots & \ddots & \vdots & \vdots \cr 
\vdots & \vdots & \vdots & \ddots & \ddots & \vdots & \vdots \cr 
0 & 0 & 0 & \cdots & \cdots & \alpha_{\beta-1} & 0\cr 
0 & 0 & 0 & \cdots & \cdots & \gamma_\beta & \alpha_\beta X\cr
\end{pmatrix},
\ee
where
\be
\alpha_k \equiv \begin{pmatrix} C_k & 1 \cr -B_k & 0 \cr \end{pmatrix};
\quad \gamma_k \equiv \begin{pmatrix} 0 & -B_k \cr 1  & -C^\dagger_k\cr \end{pmatrix};
\quad X \equiv \begin{pmatrix} 1 & 0 \cr 0 & T_3\cr \end{pmatrix};
\quad Y  \equiv \begin{pmatrix} T_3^\dagger & 0 \cr 0 & 1 \end{pmatrix}. 
\ee
Using the formula for the determinant of the above matrix
from~\cite{Neuberger:1997bg}, we arrive at
\be
\det D(M) = \left[ \prod_{j=1}^\beta \det \alpha_j \right] \det \left [ X -
  \left(\prod_{k=\beta}^1 \tc_k\right) Y\right],
\ee
where the hermitean transfer matrix $\tc_k$ associated with propagating
the fermion across the $k$-th slice is
\be
\tc_k\equiv-\alpha_k^{-1} \gamma_k =
\begin{pmatrix}
B^{-1}_k & -B^{-1}_k C^\dagger_k \cr 
-C_k B^{-1}_k  & C_k B^{-1}_k C^\dagger_k + B_k
\end{pmatrix}.
\ee
The final expression for the fermion determinant is
\be
\det D (M)=\left(\prod_{j=1}^\beta\det B_j\right)\det T_3\det\hc\qquad\text{where}\qquad
\hc\equiv
{\mathbf 1}
-\left (\prod_{k=\beta}^1\tc_k\right) T_3^\dagger
.\label{detplus}
\ee  
This is the main formula that we use repeatedly in order to understand
the phase of the  determinant, $\Gamma$, in this paper. Since $B_j$
is positive definite, the phase becomes
\be
\exp\left(i\Gamma\right)=\frac{\det D(M)}{|\det D(M)|}=\det T_3 \frac{\det \hc}{\left|\det\hc\right|}.
\label{detphs}
\ee  
If  $\xi_i$ are the  $2V$ eigenvalues  of $\prod_{k=\beta}^1\tc_kT_3^\dagger$, 
then
\be
\exp\left(i\Gamma\right)=\det T_3 \prod_{i=1}^{2V}\frac{1-\xi_i}{|1-\xi_i|}.
\ee
The positivity of the hermitean transfer matrices $\tc_k$ follow from
the positivity of $B_k$ since
\be
\begin{pmatrix} u^\dagger & v^\dagger \cr \end{pmatrix}
\tc_k \begin{pmatrix} u \cr v \cr \end{pmatrix}
= (u-C_k^\dagger v)^\dagger B_k^{-1} (u-C_k^\dagger v) + v^\dagger B_k v > 0.
\ee
In addition, they satisfy the {\sl unitarity} property $\det \tc_k = 1$.

\subsection{Free field theory}\label{sec:free}
In this subsection, we find the eigenvalues and eigenvectors of $\tc$
(we can drop the subscript $k$) for free field theory,  where all the
gauge-links are set to identity. The momenta $p_1$ and $p_2$ in the
$xy$-plane are
\be
p_i=\frac{2\pi n_i}{L}\qquad\text{where}\qquad n_i=0,1,\ldots L-1.
\ee
When expressed in this momentum basis, both $B$ and $C$ are the numbers 
\be
b = 1 - M + 2\sum_{j=1}^2 \sin^2{\frac{p_j}{2}}\qquad\text{and}\qquad c = i \sin p_1 + \sin p_2,
\ee
respectively. Thus $\tc$ becomes 
\be
\tc(n_1,n_2) = \begin{pmatrix} \frac{1}{b} & -\frac{c^*}{b} \cr -\frac{c}{b} & b+ \frac{|c|^2}{b}
\end{pmatrix}.
\label{tmom}
\ee
The eigenvalues of $\tc(n_1,n_2)$ are $e^{\pm \lambda_p }$ with
\be
\lambda_p = \cosh^{-1} \frac{1+b^2 + |c|^2}{2b}.
\ee
The corresponding normalized eigenvectors for the zero mode $(0,0)$
and the doubler modes $(\pi,0), (0,\pi)$ and $(\pi,\pi)$ are
\be
\ket{p+}= \begin{pmatrix} 1 \cr 0 \end{pmatrix}\qquad\text{and}\qquad
\ket{p-}= \begin{pmatrix} 0 \cr 1 \end{pmatrix}\qquad\text{when}\qquad b<1 ;
\ee
if $b>1$, the above $\ket{p+}$ and $\ket{p-}$ get interchanged. For
other generic modes
\be
\ket{p\pm}= \frac{1}{\sqrt{|c|^2+(1-e^{\pm \lambda_p}b)^2}}
\begin{pmatrix}  c^* \cr 1- e^{\pm \lambda_p }b \cr\end{pmatrix}.
\ee
It is straightforward to extend the free theory results to a case where
 uniform spatial torons $h_1$ and $h_2$ are present. For this, one
replaces $\tc(n_1,n_2)$ by $\tc(n_1+h_1,n_2+h_2)$.

\section{Gauge field backgrounds with uniform electric and magnetic fields}\label{sec:em}
This section is devoted to gauge-field backgrounds with constant and
 uniform electric as well as magnetic fields; non-zero $q_1$, $q_2$
and $q_3$. We first consider the case when $q_1=q_2=0$ and assume
that $h_3\ne 0$.  This is a standard example to understand the role
of large gauge transformation in the parity odd part of the induced
action~\cite{Dunne:1998qy,Deser:1997nv,Deser:1997gp,Kikukawa:1997qh} and we will show
that the results using the lattice formulation are consistent with zeta
function regularization. Next, we consider the case of  static electric
fields ($q_1\ne 0$, $q_2\ne 0$ and $q_3=0$) by reducing the problem to
a free fermion propagation with continuously changing momentum along a
closed loop in a two-dimensional momentum torus. Apart from providing
a different perspective to the constant magnetic field case, this also
leads to an understanding of a parity even phase $\pi q_1 q_2$. The last
subsection deals with a numerical study of the  general case where both
the electric and magnetic fields are present.

\subsection{Uniform and static magnetic field}\label{sec:constflux}

Let us consider a gauge-field background with only a uniform magnetic
field $q_3$ and the toron $h_3$.  In this case, the matrices $\tc_k=\tc$
are time independent. The eigenvalues of $\tc$ can be written as
$e^{\pm\lambda_i}$ due to its positivity.  The matrix $\hc$ for this
static case becomes
\be
\hc_{\rm st}\equiv\mathbf{1}-\tc^\beta e^{-i 2\pi h_3},
\ee
and its
eigenvalues $1-\xi^\pm_i$ are given by
\be
\xi_i^{\pm}=e^{\pm\beta\lambda_i^\pm-i2\pi h_3}.
\ee
It is known~\cite{Narayanan:1993ss} that $q_3$ eigenvalues of $\tc$ cross
unity as a function of mass when a non-zero topological charge $q_3$
is present; when $m>0$, there are $V+q_3$ eigenvalues $e^{\lambda^+}$,
and $V-q_3$ eigenvalues $e^{-\lambda^-}$ with $\lambda^{\pm} > 0$. The
determinant of $\hc_{\rm st}$ expressed in terms of these eigenvalues is
\be
\det\hc_{\rm st}=\prod_{i=1}^{V+q_3}\left(1-e^{-i 2\pi h_3+\frac{L}{t}\lambda^+_i}\right)\prod_{j=1}^{V-q_3}\left(1-e^{-i 2\pi h_3-\frac{L}{t}\lambda^-_j}\right).
\ee
Using \eqn{detplus}, the phase of the determinant is 
\be
\Gamma=\pi q_3-2\pi h_3 q_3+\sum_{i=1}^{V+q_3}{\rm Im}\log\left(1-e^{i 2\pi h_3-\frac{L}{t}\lambda^+_i}\right)+\sum_{j=1}^{V-q_3}{\rm Im}\log\left(1-e^{-i 2\pi h_3-\frac{L}{t}\lambda^-_j}\right).
\label{phasemag}
\ee
The first term $\pi q_3$ is parity even, and it could be absorbed
by changing $h_3 \to h_3 + \frac{1}{2}$. This formula is explicitly
gauge invariant under a large gauge transformation $h_3\to h_3+1$ and
is a consequence of the gauge  invariant regularization. At any finite
fermion mass, all the $L\lambda_i^\pm$ have non-zero finite continuum
limits. At zero temperature, all the exponentials vanish leaving only
the  first two terms which do not depend on the fermion mass.

\bef
\begin{center}
\includegraphics[scale=0.7]{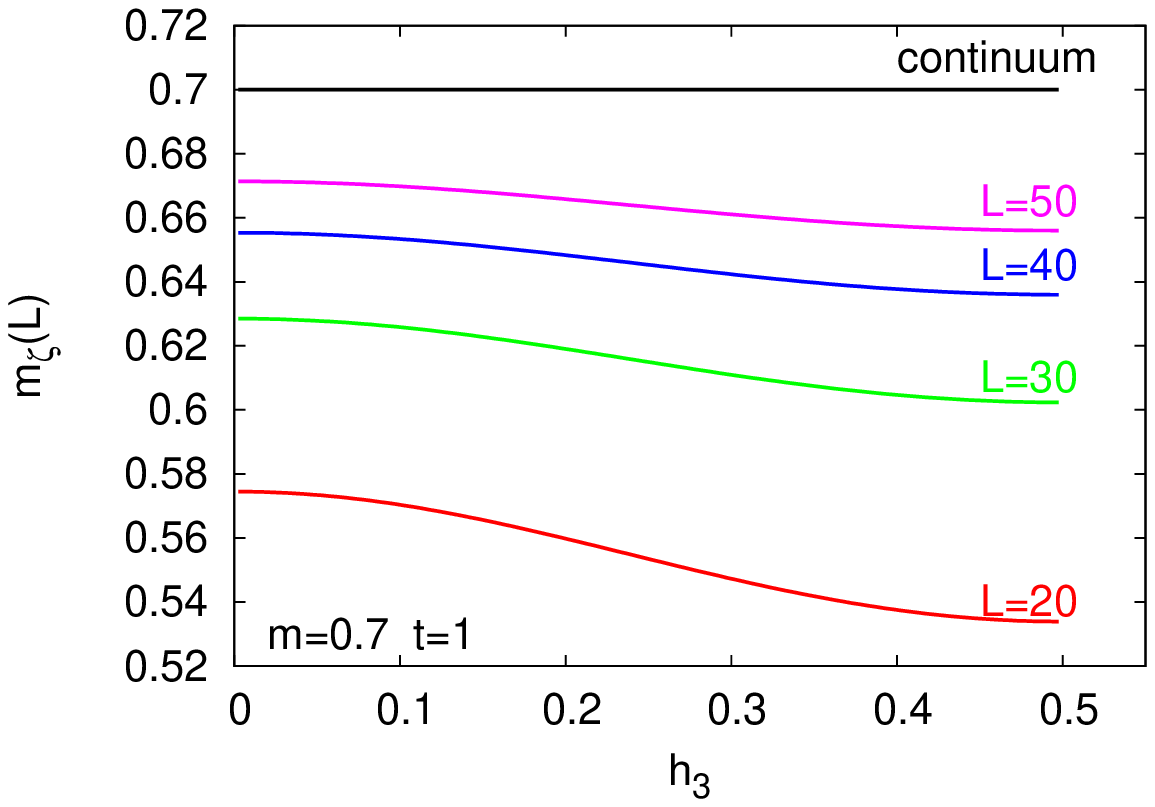}
\end{center}
\caption{
The continuum limit of $m_\zeta$ (refer \eqn{mzdef}) at various $h_3$ is shown for $m=0.7$ and 
$t=1$. The spatial lattice extent $L$ is specified on top of each curve. The continuum limit of 
$m_\zeta$ obtained by $1/L$ extrapolation is shown by the topmost black line. This continuum limit 
matches with $m=0.7$ at all $h_3$.
}
\eef{mzeta}
In order to check consistency with the results from zeta function
regularization in~\cite{Deser:1997nv,Deser:1997gp}, we computed
$\Gamma$ for several values of $m=ML$ and several values of
$t=\frac{L}{\beta}$. For the different $\Gamma$ that we computed 
at finite $L$, the mass term used in the zeta function
regularization would correspond to
\be
\frac{m_\zeta(L)}{t} = \ln\frac{\tan \Gamma}{\cos 2\pi h_3 \tan\Gamma - \sin 2\pi h_3}.
\label{mzdef}
\ee  
We extracted the continuum limit of $m_\zeta$ by fitting the results for $L=20,22,\cdots, 48,50$
using a polynomial in $\frac{1}{L}$.  We verified that the extracted
value for $m_\zeta$ matches with $m$ quite well. Our checks were made
in the range $m \in [0.2,1.0]$ and $t \in [0.1,3]$. We show an example of the 
$L$-dependence  of $m_\zeta$ in \fgn{mzeta} using $m=0.7$ and $t=1$. The continuum 
limit of $m_\zeta$ is seen to match with the $m$ used in our lattice calculation.

\subsection{Uniform and static electric fields}\label{sec:electric}

At any finite non-zero temperature and finite volume, it is  possible to
have spatially uniform and static electric fields, \ie, non-zero $q_1$
and $q_2$ which are integers. Also, $q_1$ and $q_2$ need not have the
same value.  In this subsection, we consider this case of non-zero
electric fields, but with no magnetic field. The lattice gauge field
background in \eqn{gfield} reduces to
\bea
U_1({\bf x}, k) &=&
e^{i\frac{2\pi q_2}{L \beta}k}, \cr
U_2({\bf x}, k) &=& 
e^{-i\frac{2\pi q_1}{\beta L}k}, \cr
U_3({\bf x}) &=& 
e^{-i\frac{2\pi q_2}{L}x_1+i\frac{2\pi q_1}{L}x_2}.\label{eflux}
\eea
We focus on the parity even phase arising from this configuration. At any
time-slice, the above $U_1$ and $U_2$ act like time-dependent torons $h_1$
and $h_2$ whose effect is to  offset the momentum. Switching to momentum
basis and using the replacement $n_i\to n_i+h_i$, the two dimensional
transfer matrix becomes
\be
\tc_k^{n,s} = \tc\left(n_1+q_2\frac{k}{\beta},n_2-q_1\frac{k}{\beta}\right)\delta_{n_1,s_1}\delta_{n_2,s_2},
\ee
where $\tc$ is given by \eqn{tmom} for the case with torons.  The $n$
and $s$ are the momentum indices.  The product of these matrices is
diagonal in momentum space and it is denoted as $t_{n_1,n_2}$
\be
\left[\prod_{k=\beta}^1\tc_k \right]^{n,s}\equiv\delta^{n_1,s_1}\delta^{n_2,s_2} t_{n_1,n_2}.
\ee
Since $T_3$ is already in a definite momentum $(-q_2,q_1)$, it becomes
$\mathbf{1}\delta^{n_1,s_1-q_2}\delta^{n_2,s_2+q_1}$.  Thus, 
\be
\left[\prod_{k=\beta}^1\tc_k \tc_3\right]^{n,s}= t_{n_1,n_2}\delta^{n_1-q_2,s_1}\delta^{n_2+q_1,s_2}.
\ee
We block-diagonalize the above matrix in the following way. Starting
 from an arbitrary momentum $(n_1,n_2)$, we create a cycle $\cc$ by
moving to $(n_1-q_2,n_2+q_1)$, then to $(n_1-2q_2,n_2+2q_1)$ and so on
till we are back at $(n_1,n_2)$. This will occur after $P$ steps when
both $P q_2$ and $P q_1$ become multiples of $L$. We refer to $P$ as the
cycle length and this is fixed given $q_1$, $q_2$ and $L$. The cycle $\cc$
corresponds to a $P\times P$ block, and it has  the following structure
\be
\left[\prod_{k=\beta}^1\tc_k \tc_3\right]_\cc=\begin{pmatrix} 0 &  t_{n_1,n_2} & 0 & 0 &\ldots & 0\cr
                                                0 &       0     & t_{n_1+q_2,n_1-q_1} & 0 & \ldots & 0\cr
                                                0 &       0     &  0                  & t_{n_1+2q_2,n_1-2q_1} & \ldots & 0\cr
                                               \vdots & \vdots  & \vdots             & \vdots &\ldots&\vdots\cr
                                               t_{n_1+(P-1)q_2,n_2-(P-1)q_1}& 0 & 0 & 0 & \ldots &0\end{pmatrix}.
\ee
The full momentum space will be split into several such $P\times P$
blocks. If we choose another $(n_1,n_2)$ that occurs in the above block as
the initial points of the cycle, it will only permute the entries of the
block and it will not change the determinant. Thus the full determinant
of $\hc$ factorizes into cycles with the factor from each cycle being
\be
\left[ \det \hc \right]_\cc =
\det \left [ 1 -  \prod_{r=0}^{P-1} t_{n_1-rq_2,n_2+rq_1}\right],
\label{eveqn}
\ee
which after cyclic permutation of the product of matrices becomes
\be
\left[ \det \hc \right]_\cc =
\det \left [ 1 -  \prod_{k=0}^{\beta P} \tc\left(n_1-q_2\frac{k}{\beta},n_2+q_1\frac{k}{\beta}\right) \right].
\label{cycprod}
\ee
Since
products of $\tc$ have unit determinant, it follows that
\be
\left[ \det \hc \right]_\cc =(1-\rho)\left(1-\frac{1}{\rho}\right),
\ee
where $\rho$ is the complex eigenvalue of the product of $\tc$ around
a cycle. The eigenvalues $\xi$ of the full transfer matrix are the
$P$-th roots of $\rho$ and $1/\rho$ in all the cycles.   The
complex number $\rho$ characterizes the propagation along a cycle. We
will show that there are {\sl real} cycles where $\rho$ is either real
or a complex number with unit magnitude. If the eigenvalue in the real
cycle switches sign as a function of mass, then it will be associated
with a non-zero contribution to the parity even part of $\Gamma$.

The product of $\tc$ taken along a cycle $\cc$ on the two dimensional
momentum torus has the following interpretation.  We start with
some point $(\frac{n_1}{L},\frac{n_2}{L})$ on the continuum momentum
torus of size $1\times 1$.  We move continuously along the direction
$(-q_2,q_1)$ and compute the fermion propagation along a closed loop in
this direction. One can formally convert this into an interpretation in
the continuum without worrying about regularization. The integer momenta
$(n_1,n_2)$ cover the entire range of integers in the continuum. The
continuum Hamiltonian in the $\tau$ direction at a fixed $(n_1,n_2)$ is
\be
\feyn{\tilde H}(\tau) = -\sigma_2 \frac{2\pi }{l} \left(n_1 - q_2 \tau t\right) + 
\sigma_1 \frac{2\pi }{l} \left( n_2 +q_1 \tau t\right) + \sigma_3 m.
\ee
We define fermion propagation as
\be
\tilde \phi(\tau+d \tau) = e^{\feyn{\tilde H}(\tau) d\tau} \tilde \phi(\tau),
\ee
in the limit of $d\tau\to 0$.
Let $\tilde \phi^{\pm}(\infty)$ be the result of propagation from the vector $(1,0)^t$ and $(0,1)^t$ respectively at $\tau=-\infty$.
Then,
\be
\left[ \det \hc \right]_\cc = \det \left[  1 - \begin{pmatrix} \tilde\phi^+(\infty) & \tilde\phi^-(\infty) \end{pmatrix} \right ].
\ee
in the continuum.

In order to classify cycles, consider the momenta $(n_1,n_2)$ and
$(L-n_1,L-n_2)$. From the expression for $\tc$ in \eqn{tmom},
\be
\tc\left(L-n_1-q_2\frac{k}{\beta},L-n_2+q_1\frac{k}{\beta}\right) =
\sigma_3 \tc\left(n_1+q_2\frac{k}{\beta},n_2-q_1\frac{k}{\beta}\right) \sigma_3.
\ee
Using this identity, we now show that if $\rho$ is associated with the $(n_1,n_2)$ cycle, $\rho^*$
is associated with the $(L-n_1,L-n_2)$ cycle.
Inserting $\sigma_3^2$ between the $\tc$'s in \eqn{cycprod}, we have
\be
\det \left [ 1 -  \prod_{k=0}^{\beta P} \tc\left(L-n_1-q_2\frac{k}{\beta},L-n_2+q_1\frac{k}{\beta}\right) \right]=
\det \left [ 1 -  \prod_{k=0}^{\beta P} \tc\left(n_1+q_2\frac{k}{\beta},n_2-q_1\frac{k}{\beta}\right) \right].
\ee
If we take the complex conjugate of the right hand side, then the product
becomes a product of $\tc^\dagger$ with the ordering  reversed. Using
the fact that $\tc$ are hermitean, and after changing the variable $k$
to $\beta P-k$ so as to recover the original ordering  (modulo $L$),
we obtain
\be
\det \left [ 1 -  \prod_{k=0}^{\beta P} \tc\left(L-n_1-q_2\frac{k}{\beta},L-n_2+q_1\frac{k}{\beta}\right) \right]=
\left\{ \det \left [ 1 -  \prod_{k=0}^{\beta P} \tc\left(n_1-q_2\frac{k}{\beta},n_2+q_1\frac{k}{\beta}\right) \right]
\right\}^*.
\ee
This completes the proof.

\bef
\begin{center}
\includegraphics[scale=0.4]{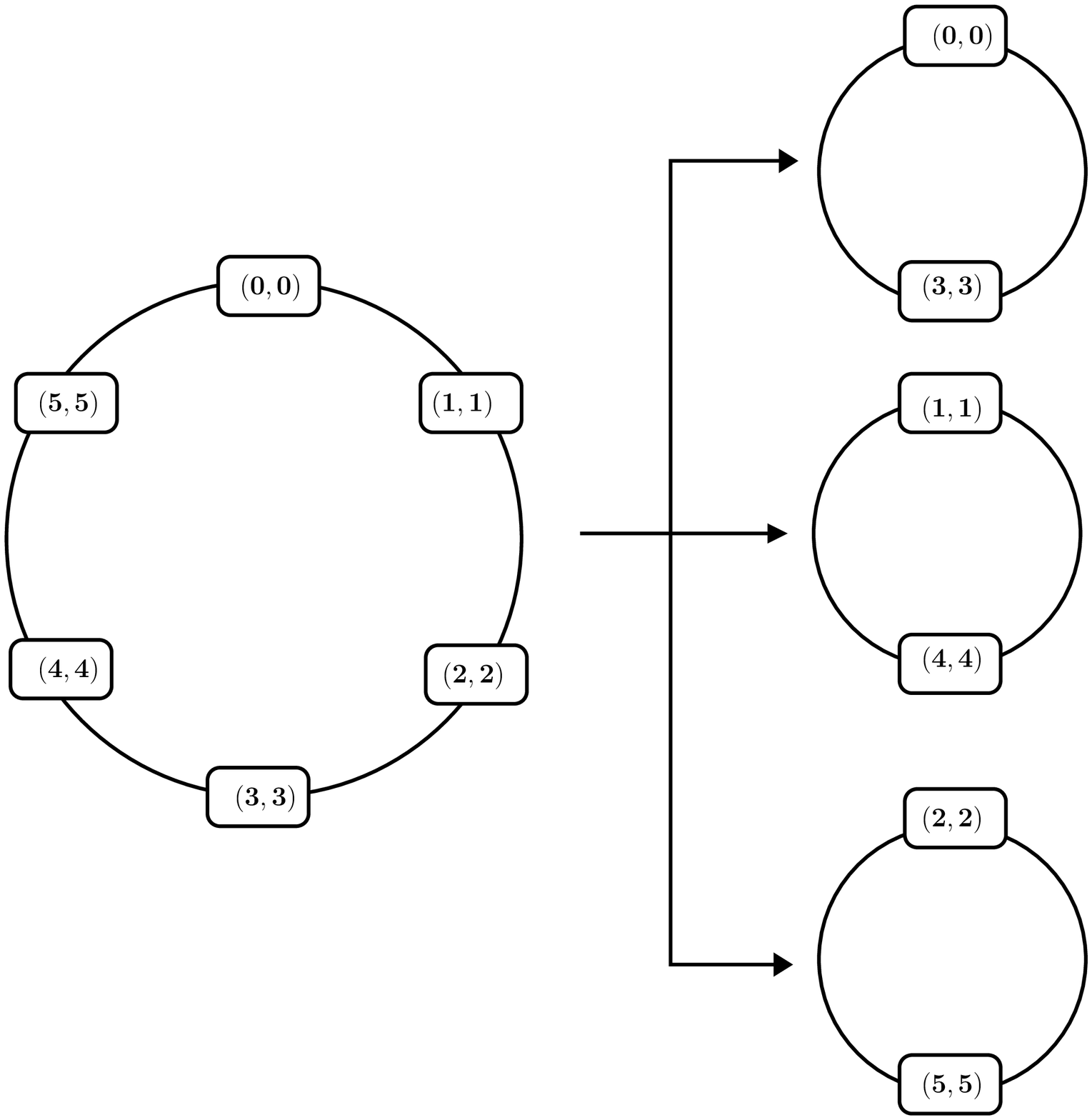}
\end{center}
\caption{On the left is the real cycle $\rc_0$ with coprime steps $q_1=q_2=1$
on a $6^3$ lattice.  This cycle has a length $P=6$.
When $q_1=q_2=3$, the cycle $\rc_0$
splits into 3 cycles each of  length $P=2$.}
\eef{split}

We classify cycles in the following way. The cycles $\cc$ and $\cc^*$
are conjugate if $(n_1,n_2)$ and $(L-n_1,L-n_2)$ belong to  $\cc$ and
$\cc^*$ respectively. If $\cc$ and $\cc^*$ are the same cycle, then we
call it a real cycle and denote it by  $\rc$. If the cycle is real then we have
\be
\left[ \det \hc \right]_\rc =(1-\rho)\left(1-\frac{1}{\rho}\right) = (1-\rho^*)\left(1-\frac{1}{\rho^*}\right).
\ee
This implies that either $\rho$ is real or $\rho$ is a complex number
with unit magnitude.  Since the eigenvalues of conjugate cycles are
related by complex conjugation,  they can be paired together to give
a positive contribution to $\det\hc$.  Therefore, only the real cycles
contribute to the phase, which can only be $\pm 1$. The following cases
are possible for the real cycles.
\begin{enumerate} 
\item When $\rho=e^{ i\phi}$, the factor will be real and positive in the above product.
\item When $\rho$ is real and negative, the contribution will be positive.
\item When $\rho$ is real and positive, the contribution will be negative.
\end{enumerate}
Depending on the number of real cycles in the above categories, the
phase could be either $+1$ or $-1$. Let us start by assuming that $q_1$
and $q_2$ are coprimes.  All the cycles have the same length, $P$, which is
even for even $L$.  Let us assume that $(n_1,n_2)$ belongs to a real
cycle. Since $(L-n_1,L-n_2)$ belongs to the same cycle, it follows that
\be
\left(L-n_1,L-n_2\right) =  \left(n_1 - rq_2, n_2+r q_1\right) + \left(k_1 L, k_2 L\right),
\ee
for some integers $r$, $k_1$ and $k_2$.  Since $q_1$ and  $q_2$ are
coprimes and $L$ is assumed to be even, it follows that $r$ is even and
\be
\left(n_1-\frac{r}{2}q_2,n_2+\frac{r}{2}q_1\right)=\left((1- k_1) \frac{L}{2},(1-k_2)\frac{L}{2}\right).
\ee
Therefore, real cycles have to contain $(n_1,n_2)=(0,0)$,
$\left(\frac{L}{2},0\right)$, $\left(0,\frac{L}{2}\right)$ or
$\left(\frac{L}{2},\frac{L}{2}\right)$.  Let $\rc_0$ denote the
cycle that contains $(n_1,n_2)=(0,0)$.  For every $(-rq_2,rq_1)$
in this cycle there is a partner, $(rq_2,-rq_1)$, in the cycle. Only
$(0,0)$, $\left(\frac{L}{2},0\right)$, $\left(0,\frac{L}{2}\right)$
or $\left(\frac{L}{2},\frac{L}{2}\right)$ have themselves as their
partner. Since each cycle has an even number of points, we conclude
that one of   $\left(\frac{L}{2},0\right)$, $\left(0,\frac{L}{2}\right)$
or $\left(\frac{L}{2},\frac{L}{2}\right)$also belongs to $\rc_0$. Since
the length $P$ can have only one factor of $2$, the number of cycles,
$\frac{L^2}{P}$ has to be even. Since the complex cycles pair up, the two
left over from $\left(\frac{L}{2},0\right)$, $\left(0,\frac{L}{2}\right)$
and $\left(\frac{L}{2},\frac{L}{2}\right)$ have to pair up and belong
to another real cycle, which we call $\rc_\pi$.  If $q_1$ and $q_2$
have a common factor, then we will assume that we choose a set of $L$
that all have this factor while taking the continuum limit. Under
such a choice, the common factor of $q_1$, $q_2$ and $L$ can be pulled
out resulting in multiples of cycles traced using coprime steps $q_1$
and $q_2$ on a smaller spatial lattice.  For the sake of  clarity, we
demonstrate the above statement in \fgn{split}, for the case $q_1=q_2=3$
on a $6^3$ lattice.

We now numerically show that the phase for $\rc_0$ is  $0$ and $\rc_\pi$
is $\pi$ for all values of $q_1$ and $q_2$ that are coprime. This enables
us to write the continuum formula for the phase for this case as
\be
\Gamma_{\rm even} = \pi (q_1+q_2+q_1q_2),\label{phq12}
\ee
in accordance with \eqn{evenintro}.  In order to maintain numerical
stability in the computation of the product of transfer matrices in
\eqn{cycprod}, we found it useful to normalize each row separately as we
multiply and cumulate the normalization factors in a single diagonal
matrix. Using this procedure we were able to work with large $L$
and $\beta$, thereby essentially seeing the behavior of cycles in the
continuum limit. The top left panel in \fgn{fl} shows the flow of the phase
from each cycle as a function of mass when the background electric flux
is $q_1=2$ and $q_2=3$.  The flow is close to what one would see in the
continuum since the computations are on a $160^3$ lattice. The phase from
the real cycle $\rc_0$  changes from being $\pi$ for $m < m_c(L)$ to $0$
for $m > m_c(L)$ for some positive $m_c(L)$, which becomes zero in the
continuum limit as shown in the top right panel of \fgn{fl}. 
The real cycle $\rc_\pi$ has a phase of $\pi$ for all
masses. The rest of the cycles are complex and come in pairs as is evident
from the plot. The combined phase is only due to the real cycles and is
$\pi$ for all masses above $m_c(L)$. This is consistent with \eqn{phq12}.

\bef
\begin{center}
\includegraphics[scale=0.7]{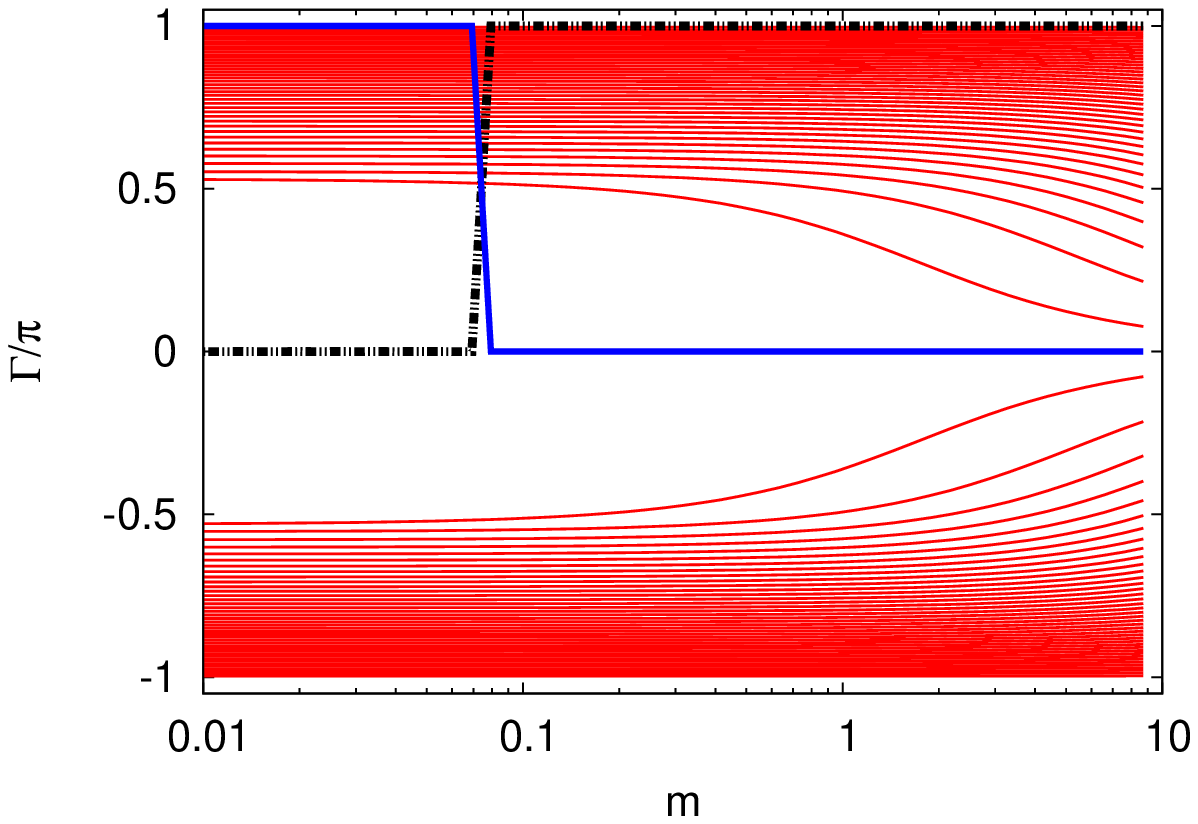}
\includegraphics[scale=0.7]{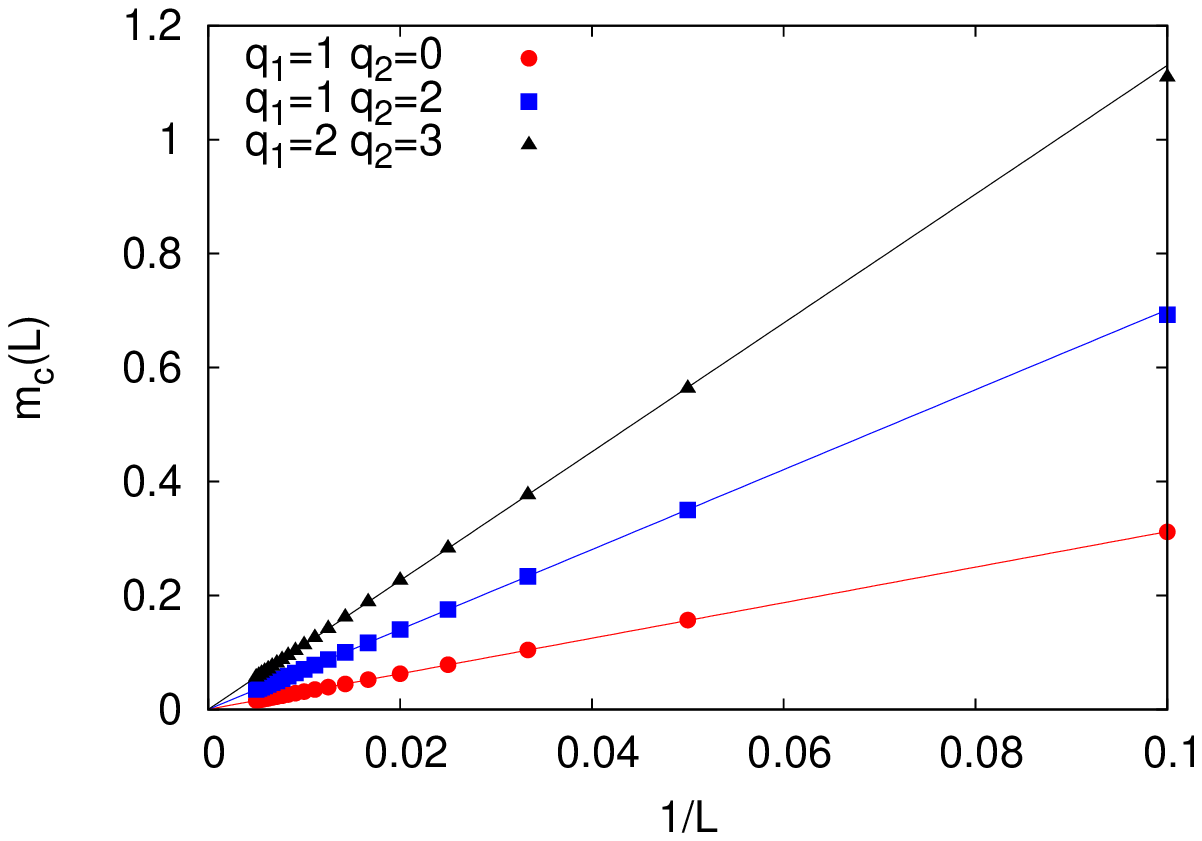}

\includegraphics[scale=0.7]{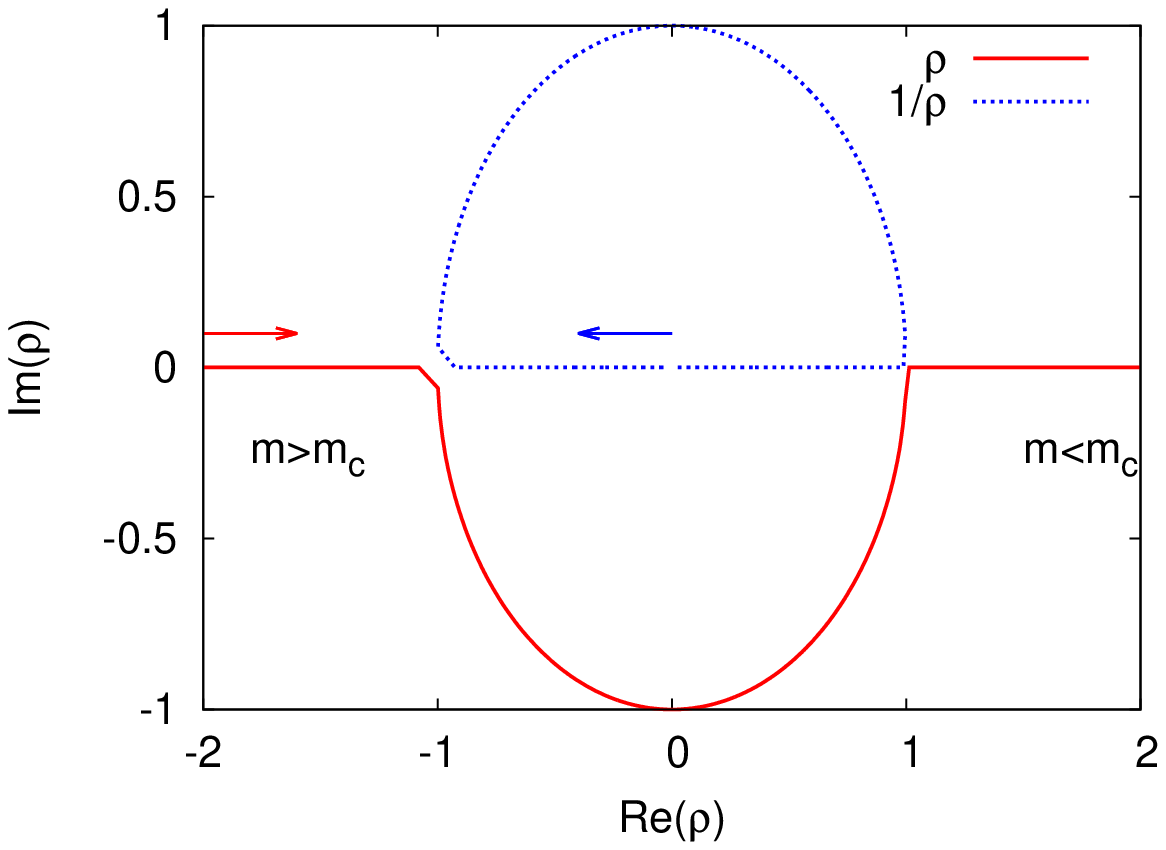}
\end{center}
\caption{ The flow of phase from each cycle as a function of mass for a
background with fixed electric flux.  \textbf{Top left panel:} The flow
for all cycles with $q_1=2$, $q_2=3$ on $160^3$ lattice is shown as red
lines.  The real cycle $\rc_0$ is shown by the solid blue line.  In addition the
plot shows the overall phase as a black dotted line. It is obvious through
a visual inspection that the eigenvalues occur as complex conjugate
pairs. The phase of the real cycle, and hence the overall phase, jump
at an $m_c$ (the points are only connected to aid the eye).  
\textbf{Top right panel:}
The lattice spacing dependence of $m_c$ is shown for various values of 
$q_1$ and $q_2$. In the continuum limit, $L\to\infty$, the
$m_c$ vanishes.
\textbf{Bottom panel:} Flow of eigenvalues of the real cycle seen on the complex
plane, in the region of crossing around $m_c$.  The plot corresponds
to flux $q_1=1$, $q_2=0$ on $4^3$ lattice in the region $0.188138<m<
0.188144$.  The pair of eigenvalues $\left(\rho,\frac{1}{\rho}\right)$
flow from real negative values for $m>m_c$ to a complex pair of unit
magnitude, and finally to real positive values.  
}
\eef{fl}

It is interesting to focus on the crossing that occurs in the $\rc_0$
cycle.  In order to zoom in on the crossing, it is better to work on
a coarse lattice and we picked $q_1=1$ and $q_2=0$ on a $4^3$ lattice
and considered $ m \in [0.188138,0.188144]$.  We look at the eigenvalue
pair $\left(\rho,\frac{1}{\rho}\right)$ as the mass is changed in this
very small range.  The flow of eigenvalues on the complex plane is
shown on the bottom panel of \fgn{fl}.  The eigenvalue pair starts out
being positive on the low end of the mass region and  approach unity at
some $m_c$ which lies within the range.  
For a range of $m$ above $m_c$, $\rho$ and
$\frac{1}{\rho}$ trace a $|\rho|=1$ locus on the complex plane. Finally,
the $\rho$ becomes real and less than zero. The background with $q_1=1$
and the one with $q_3=1$ are related by a rotation. Thus, the zero
crossing  of an eigenvalue $\lambda^+$ in the latter case, can now be
equivalently understood as the flip in  the sign of $\rho$ of the cycle
containing the zero mode. The range of $m$ where this behavior occurs
shrinks dramatically as one approaches the continuum and the value of
$m_c$ gets closer to zero.

As explained, when $q_1$ and $q_2$ are not coprimes, the cycles split into
$N$ cycles each, ($N=3$ in the example shown in \fgn{split}) depending on
the values of $q_1$ and $q_2$. Thus, all the $N$ cycles originating
from  $\rc_0$ result in a phase that switches from $\pi$ to zero as $m$
crosses $m_c$ from below.  The other $N$ cycles originating from $\rc_\pi$
always have a phase of $\pi$. Thus, the total phase becomes
\be
\Gamma=\pi (N{\rm mod\ } 2).
\ee
Only when both $q_1$ and $q_2$ are even, $N$ can be even. Thus the
expression for the phase remains as \eqn{phq12} even when $q_1$ and $q_2$
are not coprime.

Now we proceed to add $h_1$ and $h_2$ to the gauge field background
in \eqn{eflux}.  The effect is to replace $n_i$ by $n_i+h_i$ in
\eqn{cycprod}:
\be
\left[ \det \hc \right]_\cc =
\det \left [ 1 -  \prod_{k=0}^{\beta P} \tc\left(n_1+h_1-q_2\frac{k}{\beta},n_2+h_2+q_1\frac{k}{\beta}\right) \right].
\label{cycprodtor}
\ee
The full determinant still factorizes into cycles but the real cycles
now become complex and the previous complex cycles that were complex
conjugate pairs are no longer paired. If $h_1$ and $h_2$ are multiples
of $q_2/\beta$ and $q_1/\beta$ respectively, then it is possible to find
an integer $k^\prime=k-\frac{\beta h_1}{q_2}$, such that the determinant
becomes
\be
\left[ \det \hc \right]_\cc =
\det \left [ 1 -  \prod_{k^\prime=0}^{\beta P} \tc\left(n_1-q_2\frac{k^\prime}{\beta},n_2+\frac{h_1 q_1+h_2 q_2}{q_2}+q_1\frac{k^\prime}{\beta}\right) \right].
\label{cycprodtor2}
\ee
This means that at any fermion mass and temperature, the phase can only
be a function of $h_1 q_1+h_2 q_2$. In the continuum limit, the fact
that we chose a rational $h_1$ and $h_2$ should not matter. We proceed to
compute the phase per cycle and the total parity odd part of the phase of
the determinant numerically in order to understand the term in the phase
that couples $h_i$ with $q_i$. Two sample cases studied are plotted
in \fgn{flh}. Consider the case of $q_1=1$ and $q_2=0$ with $h_1=0.23$
and $h_2=0$ shown on the left panel of \fgn{flh}.  This is just a rotated
version of the case with constant magnetic flux and a temporal toron.
After removing a factor of $-1$ from the determinant due to the parity
even part of the phase, the parity odd part of the phase at the largest
mass is consistent with $-2\pi h_1 q_1$ as expected from \eqn{phasemag}.
Next, we consider the more interesting case of $q_1=2$, $q_2=3$ with
$h_1=0.37$ and $h_2=0.23$ shown on the right panel of \fgn{flh}. The
parity even part of the phase is again $\pi$ as in the previous case. The
parity odd part of the phase at the largest mass is consistent with
\be
\Gamma = - 2\pi (h_1q_1 + h_2q_2),
\label{infmass}
\ee
which is indicated by arrows in the plots.

\bef
\begin{center}
\includegraphics[scale=0.7]{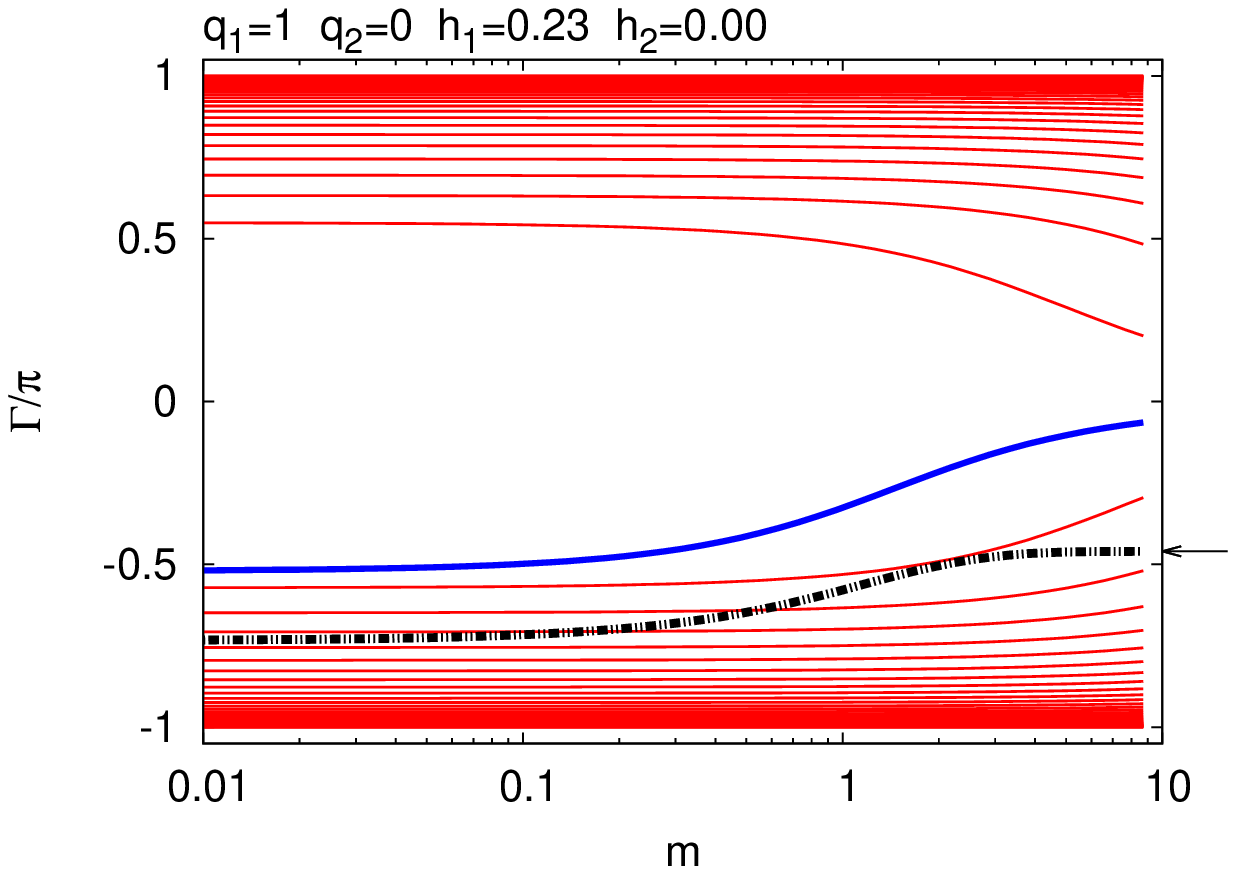}
\includegraphics[scale=0.7]{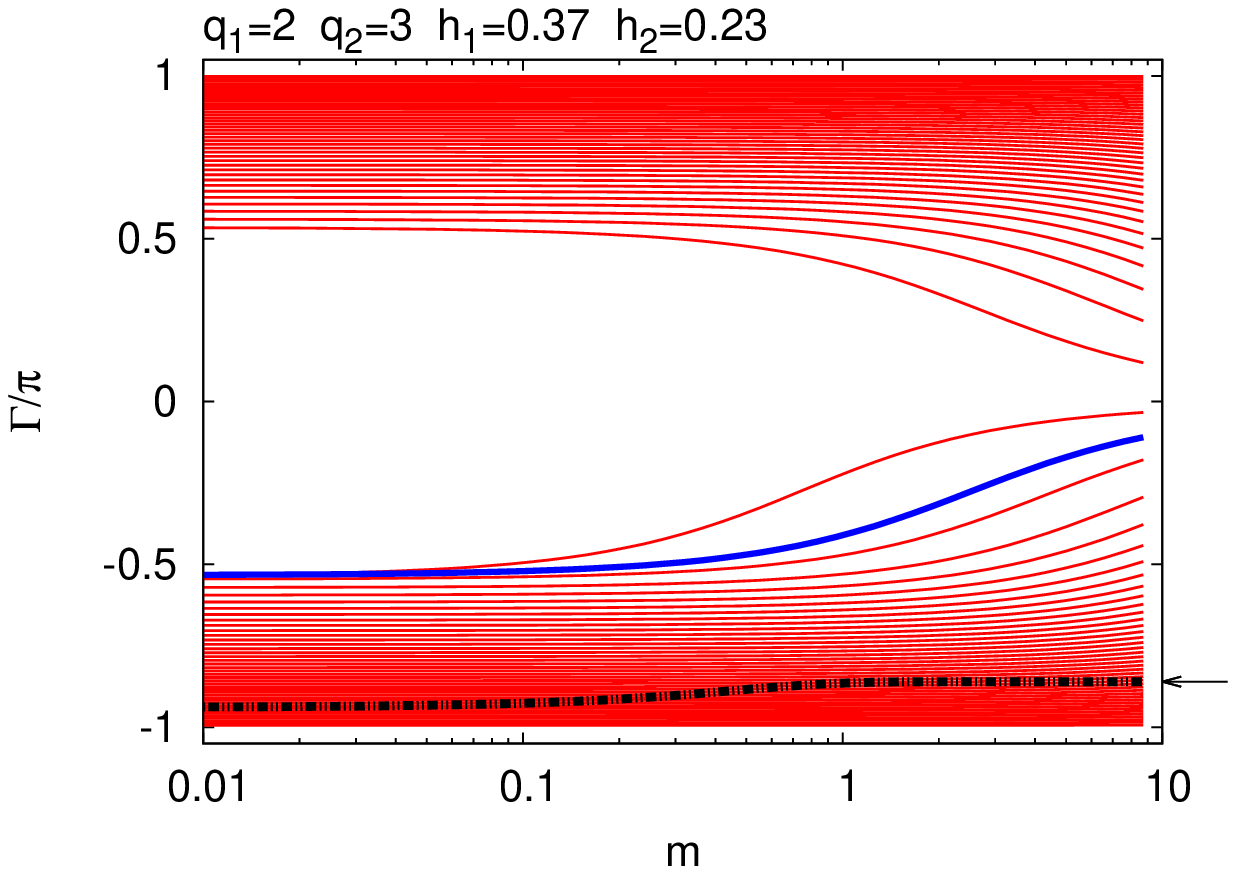}
\end{center}
\caption{
The flow of phase from all the cycles (red lines) as a function of mass
on a background with fixed electric fluxes and torons.  The left panel
shows this flow for a background with flux $q_1=1$, $q_2=0$ and torons
$h_1=0.23$, $h_2=0$ on $160^3$ lattice.  Similarly, the right panel is for
a background with flux $q_1=2$, $q_2=3$ and torons $h_1=0.37$, $h_2=0.23$.
The cycle that would have been the real cycle $\rc_0$ when torons are
absent, is specially shown by the solid blue line.  The overall parity odd
part of the phase as black dashed line.  The arrows indicate the expectations
from \eqn{infmass} for the infinite mass limit.
}
\eef{flh}

\subsection{Uniform and static electric and magnetic fields}

\bet
\begin{center}
\begin{tabular}{|c|c|c|c|}
\hline
$\quad q_1 \quad$ & $\quad q_2 \quad$ & $\quad q_3 \quad$ & $\quad \Gamma/\pi \quad$ \\
\hline
0 & 0 & 0 & 0 \\
0 & 0 & 1 & 1 \\
0 & 1 & 0 & 1 \\
0 & 1 & 1 & 1 \\
1 & 0 & 0 & 1 \\
1 & 0 & 1 & 1 \\
1 & 1 & 0 & 1 \\
1 & 1 & 1 & 0 \\
\hline
\end{tabular}
\end{center}
\caption{Phase $\Gamma$ for various combinations of $q_1$, $q_2$
and $q_3$. The ``0" represents even integers and ``1" represents odd
integers.}
\eet{bool}

Now we consider the gauge field background where electric as well as
magnetic fields are present \ie, $q_1$, $q_2$ and $q_3$ terms  are all
present in \eqn{gfield} with no torons. We are unable to study this
case analytically. Therefore, we study this general case by directly
evaluating \eqn{detplus}. We check for any  loss of precision by comparing
the determinant of the product of $\tc_k$ to 1.  Doing so, we were able
to use up to $12^3$ lattices. We find that $\det D$ is real
 for any $q_1$, $q_2$ and $q_3$. Thus, the phase can only be $\pm 1$
and its expression must be of the form
\be
\Gamma=\eta_1\pi\left(q_1+q_2+q_3\right)+\eta_2\pi\left(q_1q_2+q_2q_3+q_3q_1\right)+\eta_3\pi q_1q_2q_3,
\label{pgeven}
\ee
where $\eta_i=0$ or 1. Rotational symmetry guarantees that each
$\eta_i$ are the same for all directions. From the last section,  we
know that $\eta_1=\eta_2=1$. We do not have any analytical argument
about $\eta_3$. In \tbn{bool}, we collect our observations  about the
dependence of phase on $q_1$, $q_2$ and $q_3$. The results of the table
are robust and found to be the  same on $L=$4, 6, 8, 10 and 12 lattices,
and for various even and odd values for the $q$'s. The entries with
$q_3=0$ reiterate our observations of the last subsection. The entry
with even $q_1$ and $q_2$ includes the case $q_1=q_2=0$, which we 
understand as due to the mismatch between the number of positive and
negative eigenvalues of a  two dimensional Dirac operator. The other
cases do not offer a simple analytical explanation.  The important entry
is the last one where all $q$'s are odd. Since the phase is  even, it
implies that $\eta_3=0$. Thus, $\det D$ has a parity even phase given by
\be
\Gamma=\pi\left(q_1+q_2+q_3\right)+\pi\left(q_1q_2+q_2q_3+q_3q_1\right).
\label{peven}
\ee

\section{Perturbation theory: Parity odd contributions}\label{sec:pert}

In this section, we return to the case of uniform and static magnetic
field in the presence of a uniform toron in the Euclidean time direction
that was studied in \scn{constflux} and consider perturbations  $A_1^p$
and $A_2^p$ on this background.  We expand the determinant for $\hc$
in \eqn{detplus} in powers of  $A^p_i$ while considering the constant
flux background and the toron to be non-perturbative. The transfer matrix,
$\tc_k$, can be expanded to second order in $A_i^p$ as 
\be
\tc_k = \tc+\pac_k+\pbc_k.
\label{pertf}
\ee
The detailed expressions for $\pac_k$ and $\pbc_k$ are given in
\apx{pertap}.  We write
\be
\prod_{k=\beta}^1 \tc_k = \tc^\beta\left[1+P\right],
\ee
where
\be
P=\sum_{k=0}^{\beta-1} \tc^{-k -1} \pac_{k+1} \tc ^k
+ \sum_{k=0}^{\beta-1} \tc^{-k -1} \pbc_{k+1} \tc ^k
+ \sum_{k=1}^{\beta-1}\sum_{l=0}^{k-1} \tc^{-k-1} \pac_{k+1} \tc ^{k-l-1} \pac_{l+1} \tc^l.
\ee
From \eqn{detplus}, we have
\be
\log\det\hc =\log\det\hc_\st +\log\det\left[1+\hc_\st^{-1}P T_3^\dagger\right].
\ee
Using standard perturbation theory in the eigenbasis of the unperturbed $\tc$,
we arrive at 
\bea
\log\det\left[1+\hc_\st^{-1}P T_3^\dagger\right]&=&
\sum_{i\pm} \sum_{k=0}^{\beta-1} \frac{ e^{\pm (\beta-1)\lambda_i^\pm  -i2\pi h_3}}{1+e^{\pm \beta\lambda_i^\pm -i2\pi h_3}}\pac_{k+1}^{i\pm,i\pm}
+\sum_{i\pm} \sum_{k=0}^{\beta-1} \frac{ e^{\pm (\beta-1)\lambda_i^\pm   -i2\pi h_3}}{1+e^{\pm \beta\lambda_i^\pm -i2\pi h_3}}\pbc_{k+1}^{i\pm,i\pm}\cr
&&-\frac{1}{2}\sum_{i\pm} \sum_{j\pm} 
\sum_{k=0}^{\beta-1}
\frac{e^{\pm(\beta-1)\lambda_i^\pm \pm(\beta-1)\lambda_j^\pm -i4\pi h_3}}{\left(1+e^{\pm \beta\lambda_i^\pm -i2\pi h_3}\right)\left(1+e^{\pm \beta\lambda_j^\pm -i2\pi h_3}\right)}
\pac_{k+1}^{i\pm,j\pm} \pac_{k+1}^{j\pm,i\pm}\cr
&&+ \sum_{i\pm} \sum_{j\pm} 
\sum_{k=1}^{\beta-1}\sum_{l=0}^{k-1} \frac{e^{\pm(\beta -k+l-1)\lambda_i^\pm \pm (k-l-1)\lambda_j^\pm -i2\pi h_3}}{\left(1+e^{\pm \beta\lambda_i^\pm -i2\pi h_3}\right)\left(1+e^{\pm \beta\lambda_j^\pm -i2\pi h_3}\right)}\pac_{k+1}^{i\pm,j\pm} \pac_{l+1}^{j\pm,i\pm}+{\cal O}(A^3),
\label{perteq2}
\eea
where it is implicit that the summation over $i+$ runs up to $V+q_3$,
while that of $i-$ up to $V-q_3$. We use this general second order
perturbative expression to study two cases of interest.

\subsection{Zero temperature} \label{sec:pert0}

We assume that we are working away from the massless limit and therefore
$\lim_{L\to\infty} L\lambda_i^\pm$ are strictly greater than zero for
all $i$.  Since $\beta=\frac{L}{t}$, we see that in the limit $t\to 0$,
the first three terms in \eqn{perteq2} are real and  do not contribute
to the phase. The last term can be simplified as follows. Let
\be
\left|\pac_{k+1}^{i\pm,j\pm}\pac_{l+1}^{j\pm,i\pm}\right| < X^{i\pm,j\pm},
\ee
where the upper-bound $X^{i\pm,j\pm}$ is independent of $\beta$. Then,
the sum is bounded above by
\be
Y_{i\pm,j\pm}(\beta)=
\frac{ e^{\pm (\beta-1)\lambda_i^\pm  -i2\pi h_3}}{1+e^{\pm \beta\lambda_i^\pm -i2\pi h_3}}
\frac{1}{1+e^{\pm \beta\lambda_j^\pm -i2\pi h_3}}
\frac{\beta \left( 1- e^{\pm\lambda^\pm_j \mp \lambda^\pm_i}\right) + \left( 1- e^{\beta\left(\pm\lambda^\pm_j \mp \lambda^\pm_i\right)}\right)}{2 \left( \cosh\left(\pm\lambda^\pm_j \pm \lambda^\pm_i\right) -1 \right)} X^{i\pm,j\pm}.
\ee
Explicitly, $Y_{i+,\j+}$ and $Y_{i-,j-}$ vanish in the $\beta\to\infty$
limit. Therefore, we need to consider only the  products of
$\pac^{i+,j-}_{k+1}$ and $\pac^{i-,j+}_{l+1}$ terms. The phase is
\bea
\Gamma = \pi q_3 -2\pi h_3 q_3 
+\lim_{\beta\to\infty} \sum_{i=1}^{V+q_3} \sum_{j=1}^{V-q_3} e^{\left(\lambda_j^--\lambda_i^+\right)} \sum_{k=1}^{\beta-1}\sum_{l=0}^{k-1}
\left( e^{(l-k)\left(\lambda_i^++\lambda_j^-\right)} -
e^{(k-l-\beta)(\lambda_i^++\lambda_j^-)} \right)
{\rm \ Im\ }\pac^{i+,j-}_{k+1} \pac^{j-,i+}_{l+1}. 
\label{zerotp}
\eea 
The second term does not depend on $h_3$ and therefore the contribution
from the toron $h_3$ and the perturbative part are independent of each
other at this order.  If we assume $\frac{k-l}{L}$ is kept finite in
the infinite $L$ limit, then we can ignore the second factor in the
parenthesis of the second term.  The term $2\pi q_3 h_3$ is independent of
$m$ and  changes in multiples of $2\pi$ under large gauge-transformation
$h_3\to h_3+1$.  Even at zero temperature, the induced gauge action
is not of the type in \eqn{cs} if we include fields that do not vanish
at infinity.

\subsection{Finite temperature and $\mathbf{h_3=0}$}\label{sec:pertemp}

Our aim in this subsection is to study the purely perturbative
contribution to the phase in a possibly non-perturbative background
at finite temperature. Since we are focusing on terms of the type,
$A_1^p A_2^p$, we set $h_3=0$. In addition we only consider $A_1^p(k)$
and $A_2^p(k)$ that depend only on time.
When $h_3=0$, the first three terms in \eqn{perteq2} are real even at
finite $\beta$. The phase becomes
\be
\Gamma = \sum_{i\pm} \sum_{j\pm} 
\sum_{k=1}^{\beta-1}\sum_{l=0}^{k-1} \frac{e^{\pm(\beta -k+l-1)\lambda_i^\pm \pm (k-l-1)\lambda_j^\pm }}{\left(1+e^{\pm \beta\lambda_i^\pm }\right)\left(1+e^{\pm \beta\lambda_j^\pm }\right)}
{\rm Im} \pac_{k+1}^{i\pm,j\pm} \pac_{l+1}^{j\pm,i\pm}.
\label{pertex1}
\ee
After writing
\be 
\pac^{i\pm,j\pm}_k = \sum_{\mu=1}^2 \patc^{i\pm,j\pm}_\mu A_\mu^p(k),
\ee
we arrive at an expression for the phase, which is written concisely as
\be
\Gamma=-\sum_{k=1}^{\beta-1}\sum_{l=0}^{k-1}G(k-l)\left[ A^p_1(k)A^p_2(l) - A^p_2(k)A^p_1(l)\right].
\label{pertex2}
\ee
The form factor $G$ is
\be
G(k)=\sum_{(i\pm,j\pm)}e^{\mp\lambda_i^\pm\mp\lambda_j^\pm}\frac{\sinh\left[\left(\frac{\beta}{2}-k\right)(\pm\lambda_i^\pm\mp\lambda_j^\pm)\right]}{4\cosh\frac{\beta\lambda_i^\pm}{2}\cosh\frac{\beta\lambda_j^\pm}{2}} {\rm Im}\left[\left(\patc_1^{i\pm,j\pm}\right)^*\patc_2^{i\pm,j\pm}\right].
\ee
It satisfies the anti-symmetric property
\be
G(\beta -k) = -G(k).\label{gasymm}
\ee
This expression is valid for all  $q_3, h_1$ and $h_2$. For the
free case $q_3=0$ that we discussed in \scn{free}, the  form factor
simplifies to 
\be
G(k)=\sum_p \frac{\sinh\left[\left(\beta-2k\right)\lambda_p\right]}{2\cosh^2\frac{\beta\lambda_p}{2}} {\rm Im}\left[\left(\patc_1^{p+,p-}\right)^*\patc_2^{p+,p-}\right].
\label{freeform}
\ee
We derive the expressions for $\pac^{p+,p-}_\mu$ in \apx{mom}.  

\bef
\begin{center}
\includegraphics[scale=0.7]{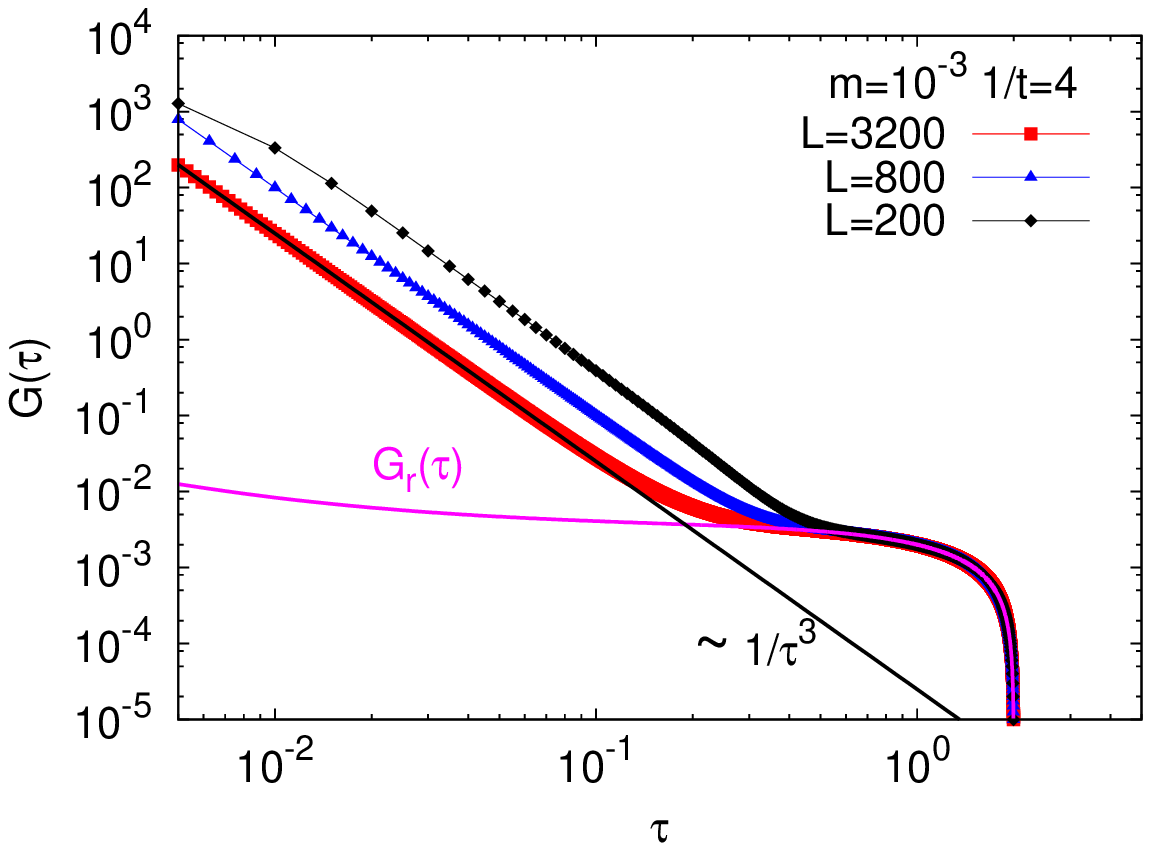}
\includegraphics[scale=0.7]{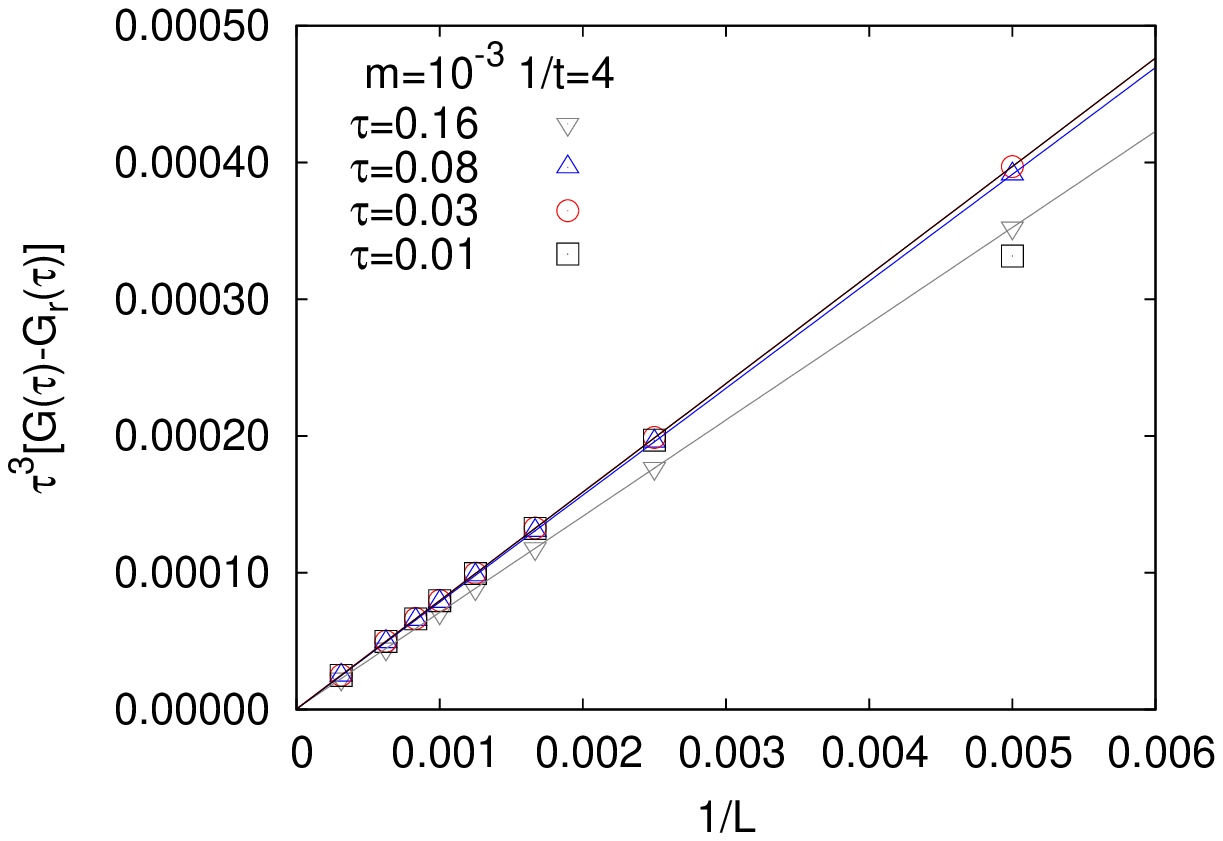}
\end{center}
\caption{
\textbf{Left panel}: The form factor $G(\tau)$ at fermion mass
$m=10^{-3}$  is shown at a temperature $t=0.25$ for the free $q_3=0$ 
background gauge-field with no torons.  The different symbols correspond 
to different $L$. At all $L$, $G(\tau)$ shows a $\tau^{-3}$ behaviour at
small values of $\tau$. The best fit for this power-law $\tau^{-3}$ using
the $L=3200$ data are shown as the black straight line.  When the continuum
is approached by increasing $L$, the power-law behaviour shifts to the
left on the log-log plot such that $G(\tau)$ approaches $G_r(\tau)$
(refer \eqn{regpart}) at all $\tau$. This continuum limit $G_r(\tau)$
is shown as the magenta curve.  \textbf{Right panel}: The approach to 
continuum of the scaled variable $\tau^{3}\left[G(\tau)-G_r(\tau)\right]$
is shown at various $\tau$.  At all $\tau$, it approaches
0 with a dominant $\frac{1}{L}$ scaling. For very small $\tau$, there is data
collapse suggesting a perfect $\tau^{-3}$ scaling. At larger $\tau$,
there are corrections to this scaling. However, it is the most singular
$\tau^{-3}/L$ behaviour of $G(\tau)$ that is important to the phase of
$\det D$.
}
\eef{div}

The behavior of this form factor $G$ is shown as a function of time
$\tau=\frac{k}{L}$ in \fgn{div}. The observations about the long and
short distance behaviour of $G(\tau)$ seen in the two panels of the
figure can be summarized in the following way. The $G(\tau)$ has a
leading $\frac{1}{L}$ lattice correction. However, the coefficient of
$\frac{1}{L}$ shows a singular $\tau^{-3}$ behaviour.  That is, the
approach to the continuum limit is given by
\be
G(\tau L) = G_r(\tau) + \frac{1}{L} G_s(\tau) + {\cal O}\left(\frac{1}{L^2}\right),
\label{divreg}
\ee
where $G_r(\tau)$ is the continuum limit, while the singular coefficient
of the dominant $\frac{1}{L}$ correction is given by
\be
G_s(\tau) = \frac{f(m,t)}{\tau^3} + {\cal O}\left(\tau^{-2}\right),
\ee
for some mass and temperature dependent function $f$.  The continuum
limit, $G_r(\tau)$, seems to be well described by the regulator
independent limit obtained by replacing $\lambda_p L$ by its $p\approx 0$
limit, $\Lambda_p$ \ie,
\be
G_r(\tau)=\sum_{n}\frac{m}{\Lambda_n} \frac{\sinh\left[\left(\frac{1}{t}-2\tau\right)\Lambda_n\right]}{2\cosh^2\left[\frac{\Lambda_n}{2t}\right]},
\label{regpart}
\ee
making use of
\be
\Lambda_n\equiv\lim_{L\to\infty}L\lambda_p = \sqrt{m^2+4\pi^2(n_1^2+n_2^2)}\qquad\text{and}\qquad\lim_{L\to\infty}{\rm Im}\left[\left(\patc_1^{p+,p-}\right)^*\patc_2^{p+,p-}\right]=\frac{m}{\Lambda_n},
\label{clambda}
\ee
for all momenta even though they only hold true for $p_i\approx 0$. The
above observations about $G$ are seen at all mass and temperature.

Let us consider the following perturbative fields chosen such that there
is a non-zero Chern-Simons action:
\be
A^p_1(k)=\frac{c}{L}\sin\left(\frac{2\pi n_3 k}{\beta}\right)\qquad\text{and}\qquad A^p_2(k)=\frac{c}{L}\cos\left(\frac{2\pi n_3 k}{\beta}\right).
\label{perta}
\ee
The phase becomes
\be
-\frac{\Gamma}{c^2}=
\lim_{L\to\infty}\frac{1}{L^2}\sum_{k=1}^{\beta-1}(\beta-k)G(k)\sin\left(\frac{2\pi n_3 k}{\beta}\right)
= 
\lim_{L\to\infty}\frac{\beta}{L^2}\sum_{k=1}^{\frac{\beta}{2}} G(k)\sin\left(\frac{2\pi n_3 k}{\beta}\right),
\ee
where we have made a change of variable from $k$ and $l$ to $k-l$, and
used the antisymmetry property of $G(k)$ in \eqn{gasymm}.  Inserting
\eqn{divreg} for $G(k)$, we obtain
\be
-\frac{\Gamma}{c^2}=
\lim_{L\to\infty}\frac{\beta}{L^2}\sum_{k=1}^{\beta-1} G_r(k)\sin\left(\frac{2\pi n_3 k}{\beta}\right)
+
\lim_{L\to\infty}\beta\sum_{k=1}^{\frac{\beta}{2}} \frac{f(m,t)}{k^3}\sin\left(\frac{2\pi n_3 k}{\beta}\right).
\ee
The first term arising from the continuum part of $G(\tau)$ can be
converted to an integral.  The second term that arises from the singular
part contributes in the continuum due to the $\tau^{-3}$ behavior. The
two terms can be expressed as
\be
-\frac{\Gamma}{c^2}=\frac{1}{t} \int_0^{\frac{1}{2t}} 
 G_r(\tau)\sin\left(2\pi n_3 \tau t\right) d\tau
+
2\pi n_3 f(m,t) \zeta(2).
\ee
The second term is proportional to the momentum $n_3$ and hence it is
indeed the local Chern-Simons term.  It contributes both in the infinite
mass and massless limit showing that the parity odd contribution is
regulator dependent~\cite{Redlich:1983dv,Coste:1989wf}.  At very low
but non-zero temperatures, the contribution from the first term behaves as
\be
\frac{\Gamma_{\rm reg}}{c^2}\approx-\frac{\pi n_3}{2}\sum_{n_1,n_2=0}^\infty\frac{m\left(1-e^{-\frac{2\Lambda_{\widetilde n}}{t}}\right)}{\Lambda_{\widetilde{n}}\left[\Lambda_{\widetilde{n}}^2+n_3^2\pi^2 t^2\right]}\qquad\text{where}\qquad \widetilde{n_i}=n_i+h_i,
\label{regamma}
\ee
after integration over $\tau$. This right away makes it explicit the
dependence of the phase on the torons $h_1$ and $h_2$ in the $q_3=0$ background.
When the torons are absent, this infinite sum suffers from
an infra-red divergence when $t\to 0$ limit is taken before the  $m\to0$
limit. But the sum becomes zero when the two limits are interchanged.
 In the $m\to\infty$ limit, the infinite sum always vanishes.
Thus, the phase from the regular term is zero in both the infinite and zero mass limits
and only the singular part contributes to the parity odd phase in these two limits.
At any finite and non-zero mass the contribution from the regular term is not local since
it is not linear in $n_3$.

\bef
\begin{center}
\includegraphics[scale=0.7]{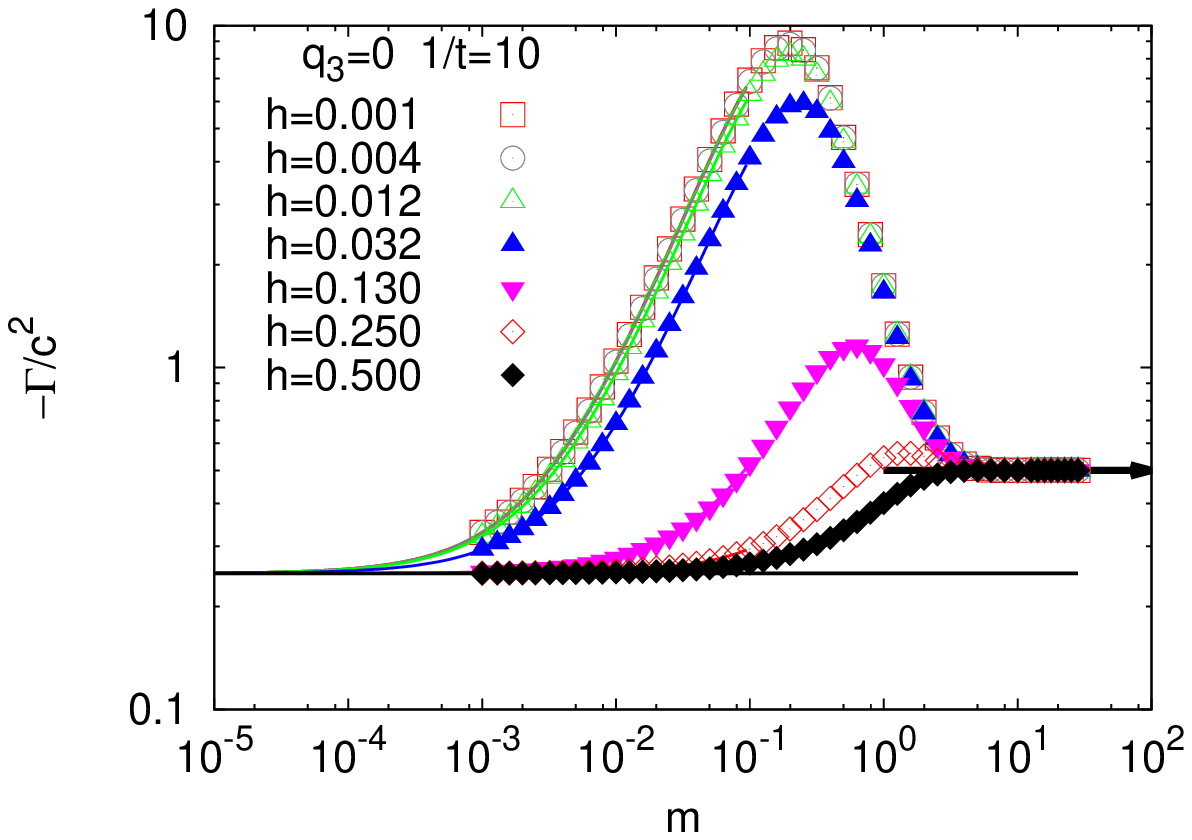}
\includegraphics[scale=0.7]{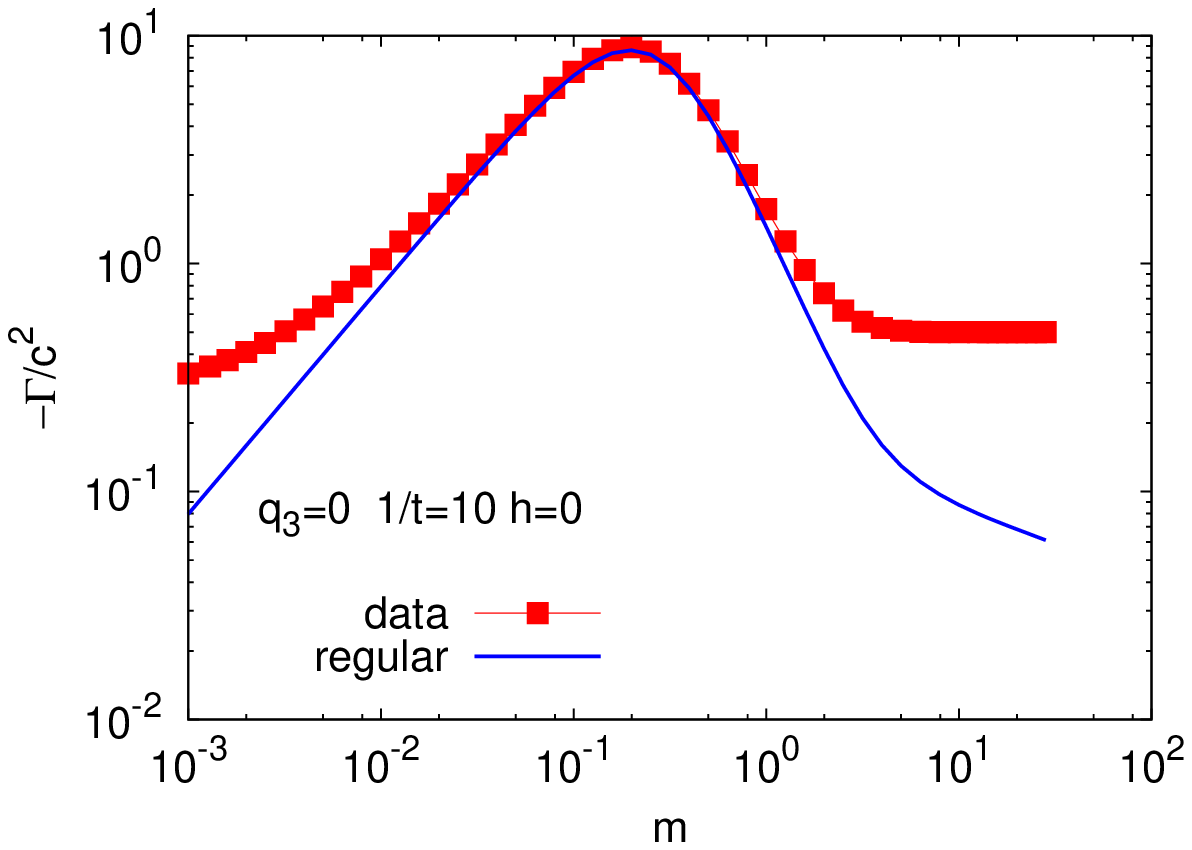}
\end{center}
\caption{ The cross-over of phase from the $m\to 0$ limit (which is
$\Gamma=-\frac{c^2}{4}$) to the $m\to\infty$ limit ($\Gamma=-\frac{c^2}{2}$) when
$q_3=0$. On the left panel, the mass dependence of $\Gamma$ is shown
for various values of $h_1=h_2=h$ specified by  different symbols. For
large values of mass ($m\gtrsim 10$), the phase is $-\frac{c^2}{2}$. For small
values of mass, the  phase approaches $-\frac{c^2}{4}$ as seen by extrapolation
(solid lines) using data points with $m<0.1$.  On the right panel,
 the phase $\Gamma$ (red squares) and the phase calculated only using
 $G_r(\tau)$ (blue line) are compared when $h=0$.  }
\eef{q0}

The above discussion shows where the parity breaking phase arises
at different masses.  We now present results on the phase directly
calculated using \eqn{pertex2}.  On the left panel of \fgn{q0}, we
show the behaviour of the phase as a function of fermion mass, for the
perturbation in \eqn{perta} on a $q_3=0$ background. We show the behavior
at various values of  $h_1=h_2=h$, and at a temperature $t=0.1$.  We did
the numerical calculation using lattices with $L=$ 60, 80, 100, 120,
140 and 160. With these, we did a continuum extrapolation for $\Gamma$
using a fourth order polynomial in $\frac{1}{L}$.  Changing the order of
the polynomial to 3 or 5 made little difference  to the results. In the
figure, we show these continuum extrapolated values.  When $m\to\infty$,
the phase becomes $-\frac{c^2}{2}$ which is consistent with a Chern-Simons
coefficient  $\kappa=-1$. Using the values of phase for $m<0.1$, we
extrapolated the results to $m=0$ using a fourth order polynomial in
$m$. These extrapolations are shown by the solid lines. The extrapolated
curves smoothly approach $-\frac{c^2}{4}$ as $m\to 0$, independent of
$h$. This corresponds to a Chern-Simons coefficient $\kappa=-\frac{1}{2}$,
which is consistent with~\cite{Coste:1989wf}.  At other intermediate
values of $m$, we find a strong dependence on $h_1$ and $h_2$, which is
expected from the above discussions for the $q_3=0$ case.  From the right
panel of \fgn{q0}, it is clear that the toron dependence of the phase
indeed comes from $G_r(\tau)$.  As $t$ becomes smaller, the peak gets
higher and shifts to smaller values of $m$ according to \eqn{regamma}.
\bef
\begin{center}
\includegraphics[scale=0.75]{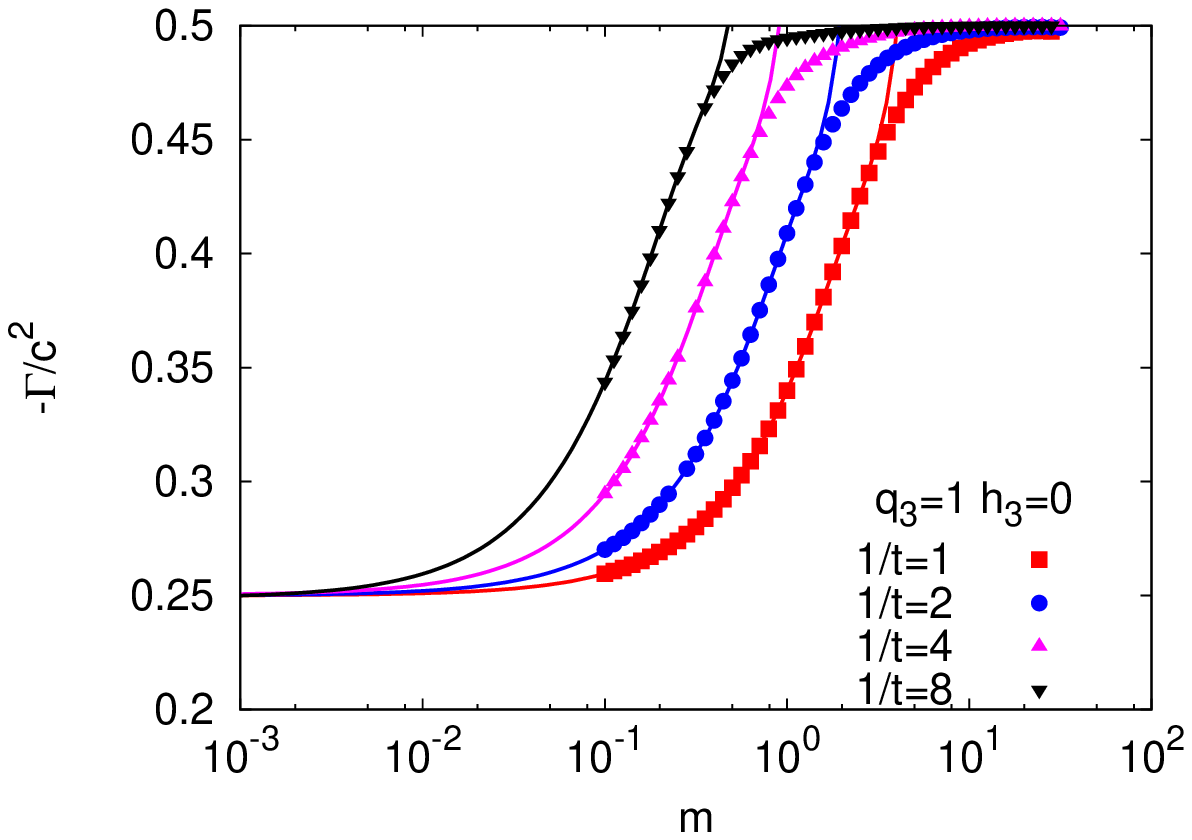}
\end{center}
\caption{
The mass dependence of phase $\Gamma$ with a flux $q_3=1$ in
the $xy$-plane. The different  symbols correspond to various temperatures
$t$ which ranges from 1 to $\frac{1}{8}$. The solid lines show the  polynomial
extrapolation of the phase from small $m<0.1$ to $m=0$. It is seen that
the phase approaches $-\frac{1}{2}$ at  large mass and the extrapolation shows
that the phase approaches $-\frac{1}{4}$ when $m\to0$.
}
\eef{q1}

In \fgn{q1}, we show a similar plot for a $q_3=1$ background. We do
not find any dependence on the spatial torons. Therefore, we show only
the  result with $h_1=h_2=0$. The different symbols are the continuum
extrapolated  results at four different temperatures. Using a similar
procedure as in the $q_3=0$ case,  we find the phase to be $-\frac{c^2}{2}$ and $-\frac{c^2}{4}$
in the infinite and zero mass limits respectively. At finite values of
$m$, there is a smooth cross-over between the two limits. At smaller $t$,
this  cross-over occurs at smaller values of $m$, well described by an
$\frac{m}{t}$ dependence of the phase.

The above mass dependence is clearly $h_1$, $h_2$ and $q_3$ dependent. 
Although implicit, one could consider them as $h_i-A^p_i$ and  $q_3-A^p_i$
terms in the  induced gauge action, that originates  from the infra-red
and would not be predicted by a pure Chern-Simons term.

\section{Conclusions}

We studied the contribution to the phase of the fermion determinant
in QED${}_3$ using lattice regularization and Wilson fermions at
finite volume and temperature. We considered non-perturbative
backgrounds that contain non-zero magnetic and electric
flux. In addition, our backgrounds also contained constant gauge
potentials referred to as torons.  In the absence of torons and
any perturbation, we studied the parity even contribution to the
phase and our result in \eqn{evenintro} is an extension of the
result in~\cite{Forte:1986em,Forte:1987kj,AlvarezGaume:1984nf}
for the case of just a magnetic flux.  In the presence of toron in
the time direction and a non-zero magnetic flux, our result using
lattice regularization agrees with one obtained using zeta function
regularization~\cite{Deser:1997gp,Deser:1997nv}. We extend this result
for the case with electric fluxes and torons. In addition to extending
the result, we provide an alternate way of understanding the parity even
contribution when one has a magnetic flux. The connection between two
dimensional topology and a parity even phase is translated to a sign
associated with the propagation of a free fermion along a closed loop
in two dimensional momentum torus where the momentum associated with
the propagation changes as one goes along the closed loop. The direction
associated with the closed loop in the two dimensional momentum torus is
proportional $(-q_2,q_1)$, the fluxes associated with the electric field.

The effect of finite temperature on the coefficient of the induced
Chern-Simons term discussed in the past~\cite{Dunne:1998qy} is addressed
here. In addition we also address the issue of finite mass. We show that
the contribution at zero mass and infinite mass only comes from the
regulator but there is also a contribution from the continuum part at
non-zero finite masses. Whereas the contribution from the regulator is
local and of the Chern-Simons type with a coefficient that is different
at zero and infinite mass~\cite{Coste:1989wf}, the contribution at any
finite non-zero mass is not local. In addition, the result depends on
the presence of torons in the space directions if there is no magnetic
flux. This is associated with the eigenvalues of the free two dimensional
Dirac operator depending on the torons and the eigenvalues of the two
dimensional Dirac operator in the presence of a non-zero magnetic flux
being independent of the torons~\cite{Sachs:1991en}.

Our studies in various non-perturbative backgrounds suggest that we can
study the following class of theories using numerical simulation:
\be
Z = \int [DU] \prod_{j=1}^{N^+} [d\psi^+_j] [d\bar\psi^+_j] \prod_{k=1}^{N^-}
[d\psi^-_k] [d\bar\psi^-_k] e^{S_g(U) + \sum_{j=1}^{N^+} \bar\psi_j \left ( \feyn D_n -B + M^+_k\right) \psi_j
\sum_{k=1}^{N^-} \bar\psi_j \left ( \feyn D_n +B - M^-_j\right) \psi_j},
\ee
with $0 < M^+_k$ and  $M^-_j < 1$.  The simplest one to simulate is the
one that does not break parity: Set $N^+=N^-=N$ and $M^-_k=M^+_k=M$.
This theory with $N$ degenerate flavors is expected to have non-zero
values for fermion bilinears that does not break parity in the
massless limit~\cite{Pisarski:1984dj}. It would be interesting to
perform a large $N$ analysis on the lattice formalism in addition to
performing a numerical simulation at small values of $N$.  Motivated
by~\cite{Witten:Simons} it would be interesting to study the theory
$N^-=0$, $N^+=N$ and $M^+_k=M$.  In particular, one could attempt
to first study this theory for large $N$ semi-classically using
the lattice formalism where the non-perturbative effects modify the
induced parity odd term at finite volume and temperature away from the
conventional Chern-Simons term in order to preserve gauge invariance. A
numerical study has to address the sign problem which might be under
control for large $N$. Since chiral symmetry is not relevant and
gauge invariance is maintained on the lattice with Wilson fermions,
numerical studies can be performed with the aim of studying massless
fermions without the necessity to use a formalism that preserves chiral
symmetry~\cite{Kikukawa:1997qh,Narayanan:1997by}.

\acknowledgments

The authors acknowledge partial support by the NSF under grant number PHY-1205396.

\appendix
\section{Expressions for $\pac$ and $\pbc$}
\label{sec:pertap}
We derive the expressions for perturbative terms $\pac$ and $\pbc$
in \eqn{pertf}. As explained in \scn{pert}, we consider  perturbative
fields $A^p_i(k)$ which are only dependent on time $k$. We expand $B_k$
and $C_k$ to second order  in perturbation theory
\bea
B_k 
    &\equiv& B +\sum_{i=1}^2 A^p_i(k) \widetilde{b}^1_i+\sum_{i=1}^2 A^p_i(k)A^p_j(k) \widetilde{b}^2_i,\cr
    &\equiv& B + b^1_k+b^2_k.
\eea
Similarly for $C_k$:
\bea
C_k &\equiv& C + \sum_{i=1}^2 A^p_i(k) \widetilde{c}^1_i+\sum_{i=1}^2 A^p_i(k)A^p_j(k) \widetilde{c}^2_i,\cr
    &\equiv& c^1_k+c^2_k.
\eea
Since, only first order terms seem to contribute to the phase, we write
down their expressions:
\be
\widetilde{b}^1_i=\frac{-i}{2}\left(T_i-T^\dagger_i\right);\quad \widetilde{c}^1_1=\frac{i}{2}\left(T_1+T^\dagger_1\right);\quad
\widetilde{c}^1_2=\frac{1}{2}\left(T_2+T^\dagger_2\right).
\label{expbc}
\ee
The $T_i$ are the forward shift operators evaluated on a free or constant
magnetic field background. Then, $B^{-1}$ can  be expanded to second
order as
\bea
B_k^{-1} &=& B^{-1} - B^{-1} b^1_k B^{-1} - B^{-1} b^2_k B^{-1} + B^{-1} b^1_k B^{-1} b^1_k B^{-1}.
\eea
Using the above expressions, one can trace the steps sketched in
\eqn{pertf} to obtain 
\bea
\tc_k &=& \tc + \pac_k + \pbc_k;\cr
\tc &=& \begin{pmatrix}
B^{-1} & -B^{-1} C^\dagger \cr 
-C B^{-1}  & C
  B^{-1} C^\dagger + B\cr 
  \end{pmatrix},\cr
  \pac_k &=&\begin{pmatrix} -B^{-1} b^1_k B^{-1} &  - B^{-1}{c^1_k}^\dagger + B^{-1} b^1_k B^{-1} C^\dagger\cr
  -c^1_k B^{-1} + C B^{-1} b^1_k B^{-1} & \Bigl( b^1_k+ c^1_k B^{-1}C^\dagger \cr
  &- C B^{-1} b^1_k B^{-1}C^\dagger
  + C B^{-1}{c^1_k}^\dagger\Bigr)\cr
\end{pmatrix},\cr
\pbc_k &=& \begin{pmatrix}- B^{-1} b^2_k B^{-1}+ B^{-1} b^1_k B^{-1} b^1_k B^{-1} 
& \Bigl(-B^{-1} {c^2_k}^\dagger+ B^{-1} b^1_k B^{-1} {c^1_k}^\dagger \cr
& +B^{-1}b_k^2 B^{-1} C^\dagger- B^{-1} b^1_k B^{-1} b^1_k B^{-1} C^\dagger\Bigr)\cr
\Bigl(-c^2_k B^{-1}+ c^1_k B^{-1} b^1_k B^{-1}&
\Bigl(b_k^2 +c_k^2 B^{-1} C^\dagger -c^1_k B^{-1} b^1_k B^{-1}C^\dagger\cr
 +C B^{-1} b_k^2 B^{-1}- C B^{-1} b^1_k B^{-1} b^1_k B^{-1} \Bigr)&+C B^{-1} b^1_k B^{-1} b^1_k B^{-1}C^\dagger+c^1_k B^{-1}{c^1_k}^\dagger \cr
&- C B^{-1} b^1_k B^{-1}{c^1_k}^\dagger + C B^{-1}{c_k^2}^\dagger\Bigr)\cr
\end{pmatrix}.\cr
&&
\eea
It is straight forward to obtain $\patc$ and $\pbtc$ from the above
expressions in terms of $\widetilde{b}^1_i$ and $\widetilde{c}^1_i$.

\section{Perturbation theory in momentum basis}\label{sec:mom}
In this appendix, we derive first order terms obtained in \apx{pertap}
in the momentum basis.  Using the Fourier transforms of \eqn{expbc},
one  obtains
\bea
\patc_i&=& \begin{pmatrix}
 \alpha_i &\beta_i \cr \beta_i^* & \gamma_i
 \end{pmatrix}\qquad\text{where},\cr
 \begin{pmatrix}
 \alpha_1 & \beta_1 \cr \beta_1^* & \gamma_1
 \end{pmatrix}
&=&
\begin{pmatrix} -\frac{\sin p_1}{b^2} & \frac{i\cos p_1}{b} + \frac{c^*\sin p_1 }{b^2} \cr
-\frac{i\cos p_1}{b} + \frac{c\sin p_1 }{b^2} & \sin p_1\left(1-\frac{|c|^2}{b^2}\right) + \frac{i(c^*-c)\cos p_1}{b} \cr
\end{pmatrix},
\cr
\begin{pmatrix}
 \alpha_2 & \beta_2 \cr \beta_2^* & \gamma_2
 \end{pmatrix}
&=&
\begin{pmatrix} -\frac{\sin p_2}{b^2} & -\frac{\cos p_2}{b} + \frac{c^*\sin p_2 }{b^2} \cr
-\frac{\cos p_2}{b} + \frac{c\sin p_2 }{b^2} & \sin p_2\left(1-\frac{|c|^2}{b^2}\right) + \frac{ (c^*+c)\cos p_2}{b} \cr
\end{pmatrix}.
\eea
Using the expressions for the eigenvalues and eigenvectors of $\tc(p)$,
\be
\patc^{p+,p-}_i = \frac{
\alpha_i |c|^2 + \beta_i c^* \left(1-e^{\lambda_p}b\right) + \beta_i^* c\left(1-e^{-\lambda_p} b\right)
+\gamma_i\left (1 + b^2 -2b\cosh\lambda_p\right)}
{\sqrt{ \left[|c|^2+(1-e^{\lambda_p}b)^2\right]\left[|c|^2+(1-e^{-\lambda_p}b)^2\right]}},
\label{egeneric}
\ee
for any generic mode. For the zero and doubler modes, it is
 \be
 \patc^{p+,p-}_i = \begin{cases} \beta_i \quad\text{if}\quad b<1 \cr
                     \beta^*_i \quad\text{if}\quad b>1.
\end{cases}
\ee
When $p\approx 0$, using \eqn{clambda}, we can replace $\lambda_p$
with $\Lambda_n/L$ in \eqn{egeneric} for $\patc_1$ and $\patc_2$. By
expanding ${\rm Im}(\patc^*_1\patc_2)$ as a power series in $1/L$,
we obtain the expression
\be
\lim_{L\to\infty}{\rm Im}\left[ \left(\patc^{p+,p-}_1\right)^*\patc^{p+,p-}_2\right]=\frac{m}{\Lambda_n}.
\label{contfp}
\ee

\bibliography{biblio}
\end{document}